\documentclass[12pt]{article}
\usepackage{latexsym}
\usepackage{amssymb}

\usepackage[dvips]{graphicx}

\newtheorem{theorem}{Theorem}
\newtheorem{lemma}{Lemma}

\newtheorem{unsolved}{Open Problem}
\newtheorem{conjecture}{Conjecture}

\newcommand{\AFigure}[3]{
  \begin{figure}
    \begin{center}
      \includegraphics{#3}
    \end{center}
  \caption{#2}\label{fig:#1}
\end{figure}}
\newcommand{\Fig}[1]{Figure \ref{fig:#1}}

\begin{document}

%\centerline{\LARGE{\bf Unsolved Problems in Visibility Graph Theory}}
\centerline{\LARGE{\bf Unsolved Problems in Visibility Graphs }}
%\centerline{\LARGE{\bf Visibility Graphs of Points, Segments}}
\bigskip
\centerline{\LARGE{\bf of Points, Segments and Polygons\footnote{The preliminary version of a part of this paper appeared in the Proceedings of India-Taiwan\\ Conference on Discrete Mathematics, Taipei, pp. 44-54, 2009.}}}
%\centerline{{\Huge{\bf Exploration in the Plane}\large{\footnote [1] {A part of the work was done when
%the first author visited Institute of Computer Science I,
%University of Bonn, Bonn 53117, Germany during April-May, 2009.}}}}

\vskip 0.7 in
\begin{center}
\mbox{\begin{minipage} [b] {2.5in}
\centerline{Subir Kumar Ghosh
%\large{\footnote{A part of the work was done when
%the author visited Institute of Computer Science, Dept. I,
%University of Bonn, Bonn 53117, Germany during April-May, 2009.}}
}
\centerline{School of Computer Science} 
\centerline{Tata Institute of Fundamental Research} 
\centerline{Mumbai 400005, India}
\centerline{ghosh@tifr.res.in}
%\centerline{March, 2007}
\end{minipage}}\hspace{0.8in}
\mbox{\begin{minipage} [b] {2.5in}
\centerline{Partha Pratim Goswami}
\centerline{Institute of Radiophysics and Electronics} 
\centerline{University of Calcutta}
\centerline{Kolkata - 700009, India}
\centerline{ppg.rpe@caluniv.ac.in}
\end{minipage}}
 
\end{center}

\vskip 0.3 in

\centerline{\bf Abstract}

In this survey paper, we present open problems and conjectures on visibility graphs of points,
 segments and
polygons along with necessary backgrounds for understanding them.
%these problems and conjectures.

%\maketitle

\section{Introduction}\label{lintroduction}
The visibility graph is a fundamental structure studied in the field of computational geometry and 
geometric graph theory, and pose some special challenges \cite{bcko-cgaa-08, g-vap-07}. 
Apart from theoretical interests, 
visibility graphs have important applications also. Some of the early applications include computing 
Euclidean shortest paths in the presence of obstacles \cite{lw-apcf-79} and decomposing 
two-dimensional shapes into clusters \cite{sh-ddsg-79}. For more on the uses of this class of 
graphs, see \cite{o-agta-87, s-rriag-92}. 

\AFigure{lvisgraph-point}{
  (a) A given set of points.
  (b) The visibility graph of the point set.
  (c) The visibility graph drawn on the point set.
% }{figs/fvisgraph-point.eps}
 }{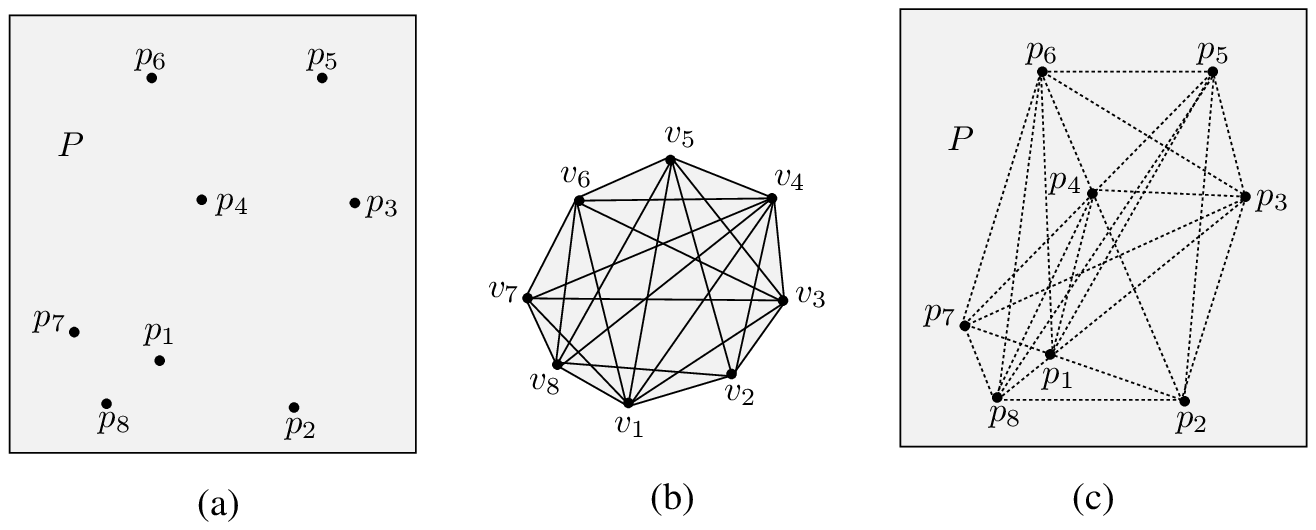}

Let $P$ be a set of $n$ points in the plane (see \Fig{lvisgraph-point}(a)).  
We say two points $p_i$ and $p_j$ of $P$ 
are \emph{mutually visible} if the line 
segment $p_ip_j$ does not contain or pass through any other point of $P$. In other words,
$p_i$  and $p_j$ are visible if $P \cap \overline{p_ip_j}=\{p_i, p_j\}$.
If a point $p_k \in P$ lies on the segment $p_ip_j$ connecting two points $p_i$ and $p_j$ in $P$, 
we say that $p_k$ blocks the visibility between $p_i$ and $p_j$, and
$p_k$ is called a {\it blocker} in $P$.  For example in \Fig{lvisgraph-point}(a), 
$p_4$ blocks the visibility between $p_2$ and $p_6$ as $p_4$ lies on the segment $p_2p_6$.

The {\it visibility graph} 
(also called the {\it point visibility graph})
$G$ of $P$ is defined 
by associating a vertex $v_i$ with each point $p_i$ of $P$ 
such that $(v_i, v_j)$ is an undirected edge of $G$ if  $p_i$ and $p_j$
are mutually visible (see \Fig{lvisgraph-point}(b)). It can be seen that if no three points of 
$P$ are collinear, i.e., there is no blocker in $P$, then $G$ is a complete graph as
each pair of points in $P$ is visible. 
Sometimes the visibility graph is drawn directly on the point set, 
as shown in \Fig{lvisgraph-point}(c).

Consider the problem of computing the visibility graph $G$ of a point set $P$. For each point
$p_i$ of $P$, sort the points of $P$ in angular order around $p_i$.  If two points $p_j$ and $p_k$ 
are adjacent in the sorted order, check whether $p_i$, $p_j$ and $p_k$ are collinear points.
By traversing the sorted order, all points of $P$, that are not visible from $p_i$, can be located
in $O(n \log n)$ time. Hence, $G$ can be computed from $P$ in $O(n ^2 \log n)$ time.
Using the result of Chazelle et al. \cite{cgl-pgd-85} or Edelsbrunner et al. 
\cite{Edelsbrunner:1986:CAL}, the running time of the algorithm can be improved to $O(n^2)$  
by computing sorted angular orders for all points together 
in $O(n^2)$ time.

\AFigure{lvisgraph-segment}{
  (a) A given set of segments.
  (b) The visibility graph of the set of segments.
  (c) The visibility graph drawn on the  set of segments.
% }{figs/fvisgraph-segment.eps}
    }{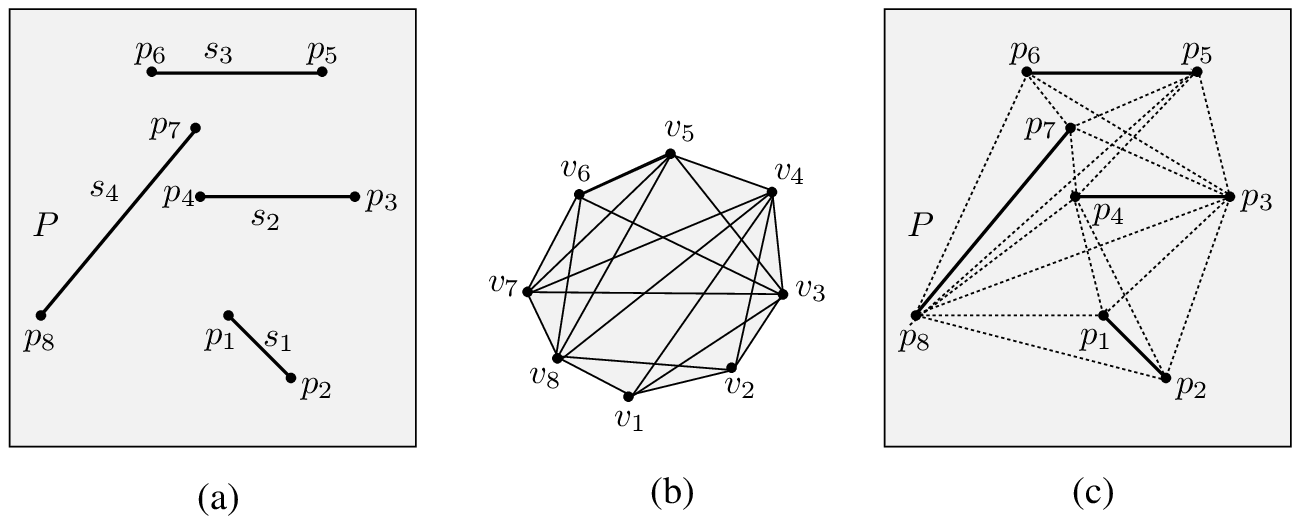}

Let $S$ be a set of $n$ disjoint line segments (see \Fig{lvisgraph-segment}(a)). 
The endpoints of segments $s_1$, $s_2, \ldots,$ $s_n$ in 
$S$ are marked as $p_1$, $p_2, \ldots,$ $p_{2n}$, where $p_{2i-1}$ and $p_{2i}$ are endpoints of 
 $s_i$. Let $P$ be the set of these endpoints $p_1$, $p_2, \ldots,$ $p_{2n}$. 
We say two points $p_i$ and $p_j$ of $P$ 
are \emph{mutually visible} if the line 
segment $p_ip_j$ does not intersect any segment $s_i$ in $S$.
This definition does not allow the segment  $p_ip_j$ to pass through another endpoint $p_k$ 
or graze along a segment in $S$.  The {\it visibility graph} 
(also called the {\it segment visibility graph or segment endpoint visibility graph})
$G$ of $S$ is defined 
by associating a vertex $v_i$ with each point $p_i$ of $P$ 
such that $(v_i, v_j)$ is an undirected edge of $G$ if  $p_i$ and $p_j$
are mutually visible (see \Fig{lvisgraph-point}(b)).
In addition, the corresponding vertices of two endpoints of every segment
in $S$ is also connected by an edge in $G$.
Sometimes the visibility graph is drawn directly on the segments, 
as shown in \Fig{lvisgraph-segment}(c).

Let $P$ be a simple polygon with or without 
holes in the plane (see \Fig{lvisgraph}(a)). We say two points $a$ and $b$ in $P$ 
are \emph{mutually visible} if the line 
segment $ab$ lies entirely within $P$. This definition allows the segment  $ab$ to pass through a 
reflex vertex or graze along a polygonal edge. The {\it visibility graph} (also called the 
{\it vertex visibility graph}) $G$ of $P$ is defined by associating a node with each vertex of $P$ 
such that $(v_i, v_j)$ is an undirected edge of $G$ if polygonal vertices $v_i$ and $v_j$ are
mutually visible. \Fig{lvisgraph}(b) shows the visibility graph of the polygon in \Fig{lvisgraph}(a). 
Sometimes the visibility graph is drawn directly on the polygon, as shown in \Fig{lvisgraph}(c). It 
can be seen that every triangulation of $P$ corresponds to a subgraph of the visibility graph of $P$. 

\AFigure{lvisgraph}{
  (a) A polygon.
  (b) The visibility graph of the polygon.
  (c) The visibility graph drawn on the polygon.
 % }{figs/fvisgraph.eps}
}{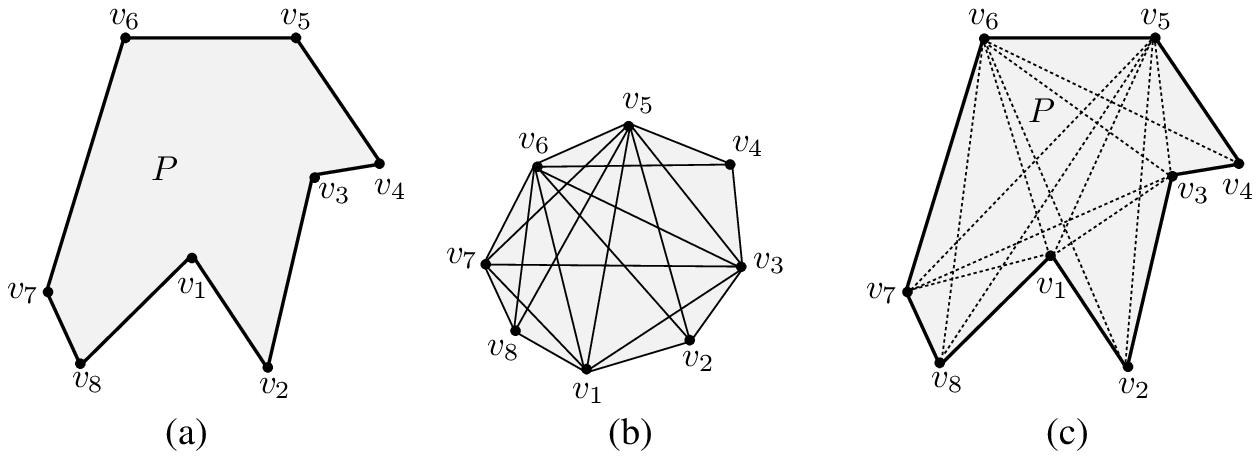}

The problem of computing the visibility graph of a polygon $P$ (with or without holes) or a set
of disjoint segments $S$
%having a total of $n$ vertices 
is well studied in computational geometry \cite{g-vap-07, lee-prp-78, lw-apcf-79, ss-spps-86}. 
Asano et al.\ 
\cite{aaghi-vdp-86} and Welzl \cite{w-cvgnl-85} proposed $O(n^2)$ time algorithms for this 
problem. Since, at its largest, a visibility graph can be of size $O(n^2)$, algorithms of 
Asano et al.\ and Welzl are worst-case optimal. The visibility graph may be much smaller than 
its worst-case size of $O(n^2)$ (in particular, it can have $O(n)$ edges) and therefore, it is 
not necessary to spend $O(n^2)$ time to compute it. Hershberger \cite{Hershberger89} developed 
an $O(E)$ time output sensitive algorithm for computing the visibility graph of a simple polygon. 
Ghosh and Mount \cite{gm-91} presented $O(n \log n + E)$ time, $O(E+n)$ space algorithm for 
computing visibility graph of polygon with holes. Keeping the same time complexity, Pocchiola 
and Vegter \cite{pv-dcg-96} improved the space complexity to $O(n)$. 

\section{Visibility graph theory: Points}

\subsection{Visibility Graphs: Recognition, Characterization, and Reconstruction}
%\subsection{Visibility Graph Recognition, Characterization, and Reconstruction}
We have stated earlier how to compute the visibility graph $G$ from a given set of
points $P$. Consider the opposite problem of determining if there is a set of points $P$ 
whose visibility graph is the given graph $G$. This problem is called the visibility graph 
{\it recognition} problem.  Identifying the set of properties satisfied by all visibility 
graphs is called the visibility graph {\it characterization} problem. The problem of 
actually drawing one such set of points $P$ whose visibility graph is the given graph $G$, 
is called the visibility graph {\it reconstruction} problem. 

\begin{unsolved}
 Given a graph $G$ in adjacency matrix form, determine whether $G$ is the visibility graph of a set
of points $P$ in the plane. 
\end{unsolved}

\begin{unsolved}
Characterize visibility graphs of point sets.
\end{unsolved}

\begin{unsolved}
Given the visibility graph $G$ of a set of points, draw the points in the plane 
whose visibility graph is 
$G$.
\end{unsolved}

%\subsection{Graph Theoretic Problems on Visibility Graphs}
%\subsubsection{Colouring Visibility Graphs}
\subsection{Colouring Visibility Graphs}

Consider the problem of colouring the visibility graph $G=(V,E)$ of a point set $P$. 
A $k$-colouring of $G$ is a function $f:V \rightarrow C$ for some set $C$ of $k$ 
colours such that $f(v) \neq f(w)$ for every edge $(v_i,v_j) \in E$. If $G$ can be coloured
by $k$ colours, $G$ is called {\it $k$-colourable}.   The {\it chromatic number} 
$\chi(G)$ is the minimum $k$ such that $G$ is $k$-colourable.
The {\it clique number} $\omega(G)$
is the maximum $m$ such $G$ contains a complete graph of $m$ vertices as a subgraph.
We start with the following lemma of K\'{a}ra et al. \cite{kpw-ocnv-2005}.

\begin{lemma}

Let $P=\{(x,y):x,y \in \mathbb{Z}\}$ be the integer lattice. Then $\chi(G)=\omega(G)=4$.
\end{lemma}

\AFigure{llattice}{
(a) The visibility graph of the integer lattice are coloured with four colours.
Only collinear edges of the visibility graph  are drawn in the figure.
(b) Five lattice points form a chordless cycle.
%}{figs/flattice.eps}
}{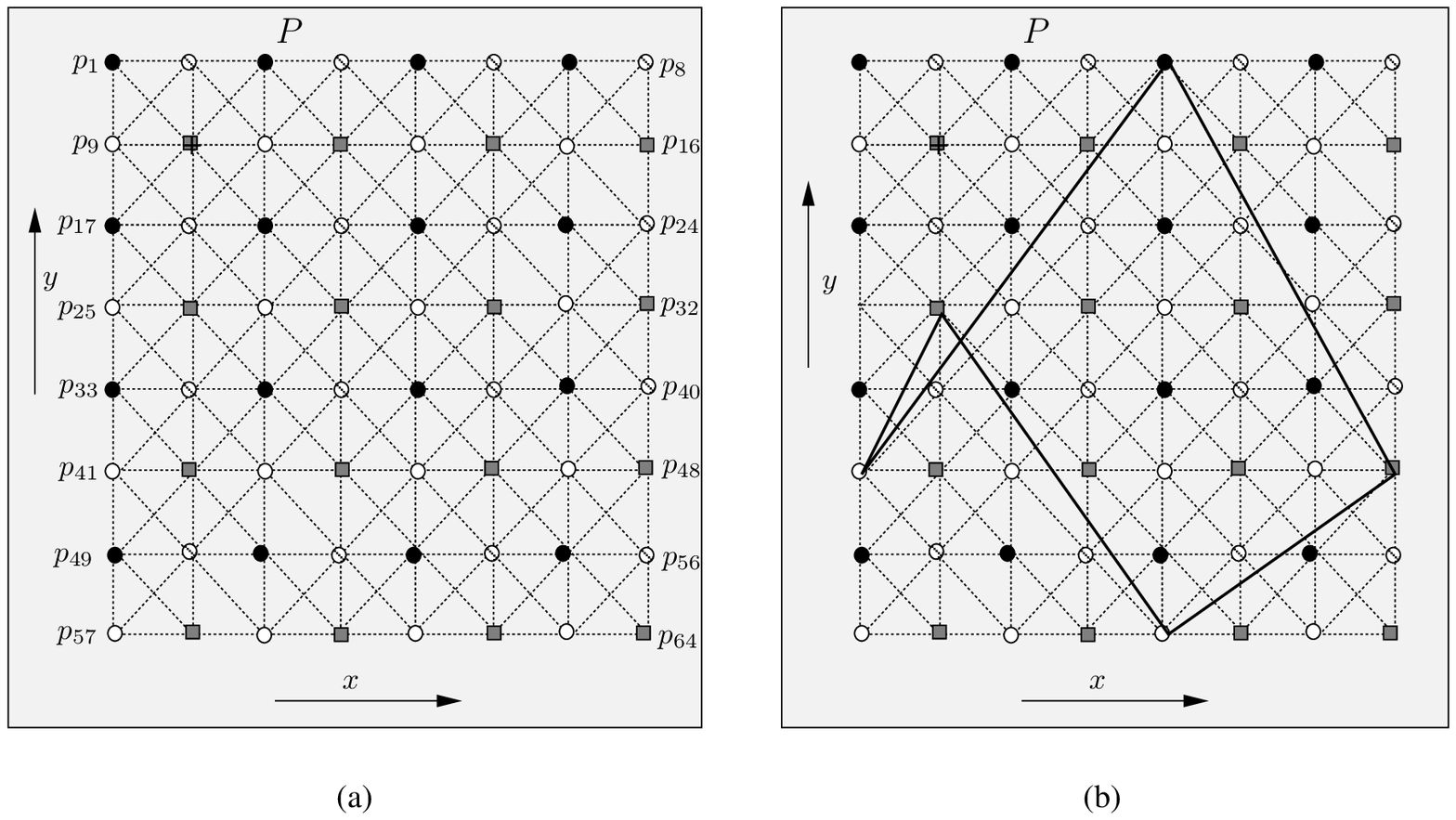}

%\Fig{llattice} shows the visibility graph of an integer lattice (all edges are not drawn).
It can be seen that all collinear lattice points on a line in \Fig{llattice}(a)
can be coloured by two colours alternatively.
Using this observation, the above lemma proves that the graph can be coloured by four colours.
They also made the following observation.

\begin{lemma}

If a point set $P \subseteq \mathbb{R}^2$ can be covered by $m$ lines, then $\chi(G) \leq 2m$.

\end{lemma}

\Fig{llattice}(a) demonstrates that 
the visibility graph of an integer 
lattice has a small chromatic
number (i.e., $4$) though the graph contains quadratic number of edges. Observe that 
the chromatic number is same as the clique number in this graph but
the graph is not a perfect graph as it contains a cycle
of five vertices without chord. For example, the five lattice 
points in $8X8$ lattice with
co-ordinates $(2,5), (1,3),(5,8),(8,3),(5,1)$ 
form a chordless cycle (see \Fig{llattice}(b)). However,
K\'{a}ra et al. felt that there is a relationship between the clique number and
chromatic number in visibility graphs of points, and made the following 
conjecture.

\begin{conjecture}\label{conject1}

There exists a function $f$ such that $\chi(G) \le f(\omega(G))$.
\end{conjecture}

\AFigure{lplanar}{ Visibility graphs of points with $\omega(G) \leq  3$ are planar.
%}{figs/fplanar.eps}
}{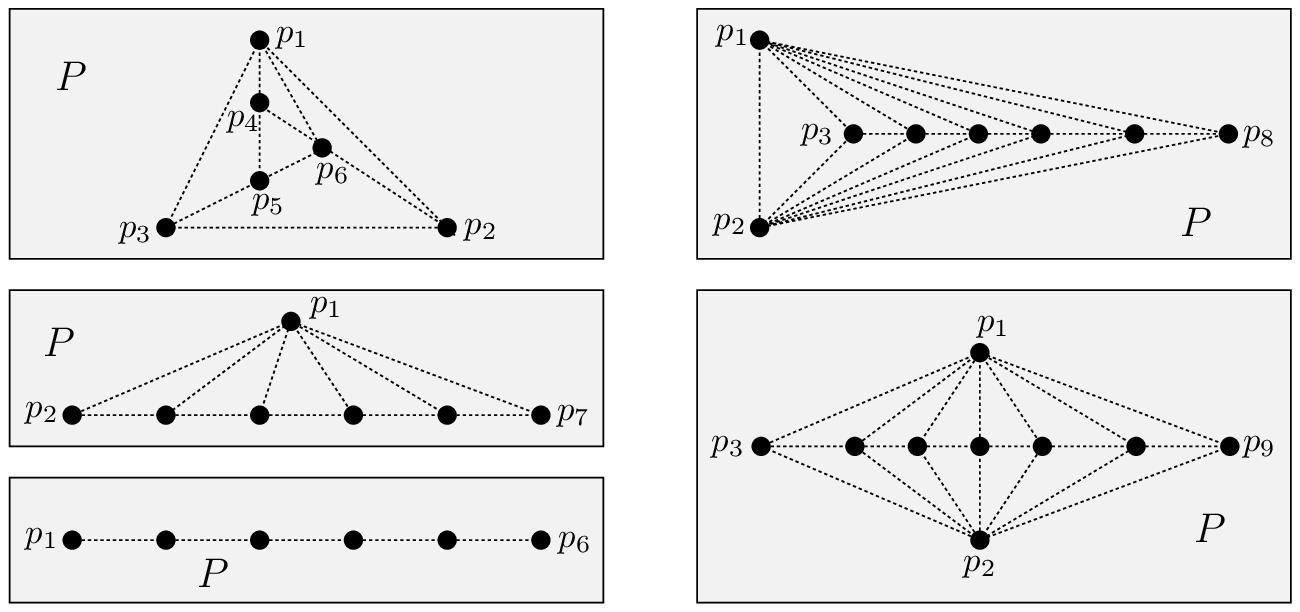}

In support of the conjecture, they proved that visibility graphs  having $\omega(G) \leq 3$ 
are planar  \cite{desw-dpgf-2007} (see \Fig{lplanar}) and they require at most 
3 colours. For  $\omega(G)= 4$,
they  showed an example (see \Fig{lconvexk} (a))  that  visibility graphs 
with $\omega(G)= 4$ require $5$ colours.

\begin{unsolved}\label{lkara1}
Prove that every visibility graph with  $\omega(G) \leq  4$ has $\chi(G) \leq 5$.
\end{unsolved}

\begin{unsolved}
Prove Conjecture \ref{conject1}  for visibility graphs with $\omega(G)=5$.
%Every visibility graph with  $\omega(G) \leq  4$ has $\chi(G) \leq 5$.
\end{unsolved}

%The conjecture is open for visibility graphs with $\omega(G)=5$. 
For visibility graphs with  $\omega(G) \geq 6$, Pfender \cite{p-vgps-2008}
proved that the conjecture does not hold as shown 
%by Pfender \cite{p-vgps-2008}.
in the following lemma.

\begin{lemma}

For every $k$, there is a finite point set $y \subset \mathbb{R}^2$ such that 
  $\chi(G) \geq k$ and $\omega(G)= 6$.

\end{lemma}

\subsection{Big Clique  in Visibility Graphs}

As stated earlier, if all points of $P$ are in general position, i.e., no three
points of $P$ are collinear,  the visibility graph $G$ of $P$ is a complete graph,
and therefore, $\omega(G)$ is the size of $P$. Consider a convex polygon formed
by points of $P$. A polygon $C$ is said to be {\it convex} if the line segment joining
any two points in $C$ lie inside $C$. Though any subset $X$ of $P$ forms
a complete graph in $G$, these points may not always form a convex polygon in $P$
(see \Fig{lconvexk}(b)). 
Even if all points of $X$ are in convex position forming a convex polygon $C$,  
some points of $P$ 
may lie inside $C$. Several papers studied these problems for point sets with or without 
collinear points. We start with the famous result of Erd\"{o}s and Szekeres \cite{es-cpg-1935}
for points of $P$ in general position.

\AFigure{lconvexk}{ 
(a) The clique number of the visibility graph is $4$ but 
the graph requires $5$ colours.
(b) Any subset of points may not form a convex polygon.
%}{figs/fconvexk.eps}
}{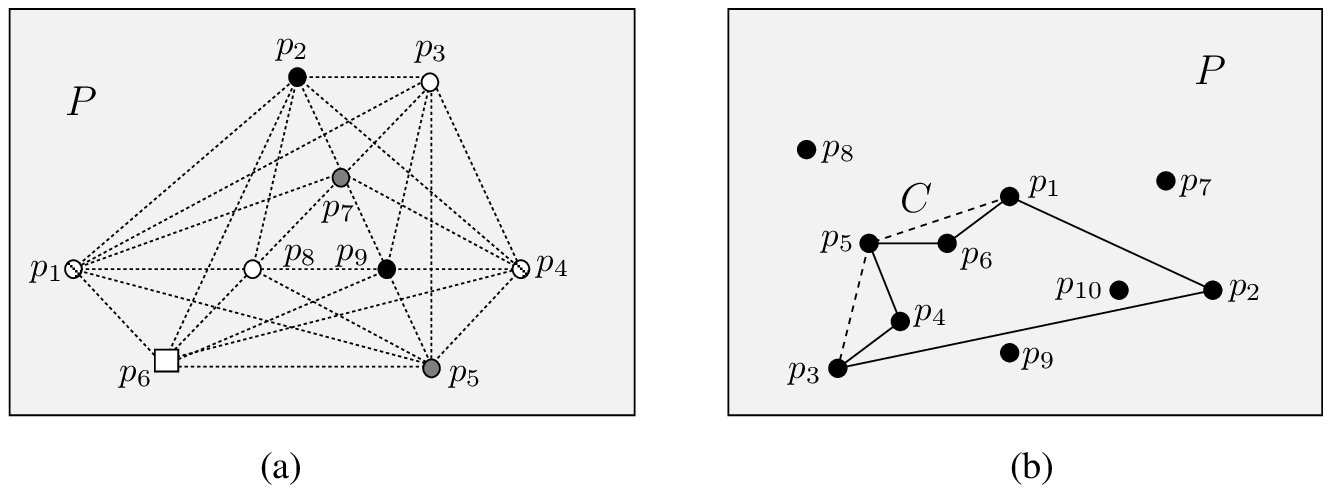}

\begin{theorem}

For every positive integer  $k$, there exists a smallest integer $g(k)$ such that any
point set $P$ of at least $g(k)$ points in general position has a subset $X$ of $k$ points 
that are the vertices of a convex polygon $C$.

\end{theorem}

It may be noted that
the existence of the value g(k) runs immediately from the famous Ramsey
theorem \cite{r-opfl-1930}.
The best known bounds for $g(k)$ are  $2^{k-2}+1 \leq g(k) \leq {{2k-5} \choose {k-2}} +2$. 
The lower and upper bounds are given by Erd\"{o}s and Szekeres
\cite{es-sepe-1961}, and  T\'{o}th and Valtr \cite{tv-estu-2005} respectively. 
For survey on this problem and many variants, see \cite{bk-pra-2001,bmp-rpd-2005,ms-espp-2000,tv-estu-2005}.

\AFigure{lconvexk1}{ 
(a) Three points are enough for an empty triangle but four
points may not always give empty quadrilateral. 
%For an empty triangle, $h(3)=3$ and for empty quadrilateral,
%$h(4)=5$.
(b) In this set of $9$ points, no subset of $5$ points forms an empty convex polygon.
%}{figs/fconvexk1.eps}
}{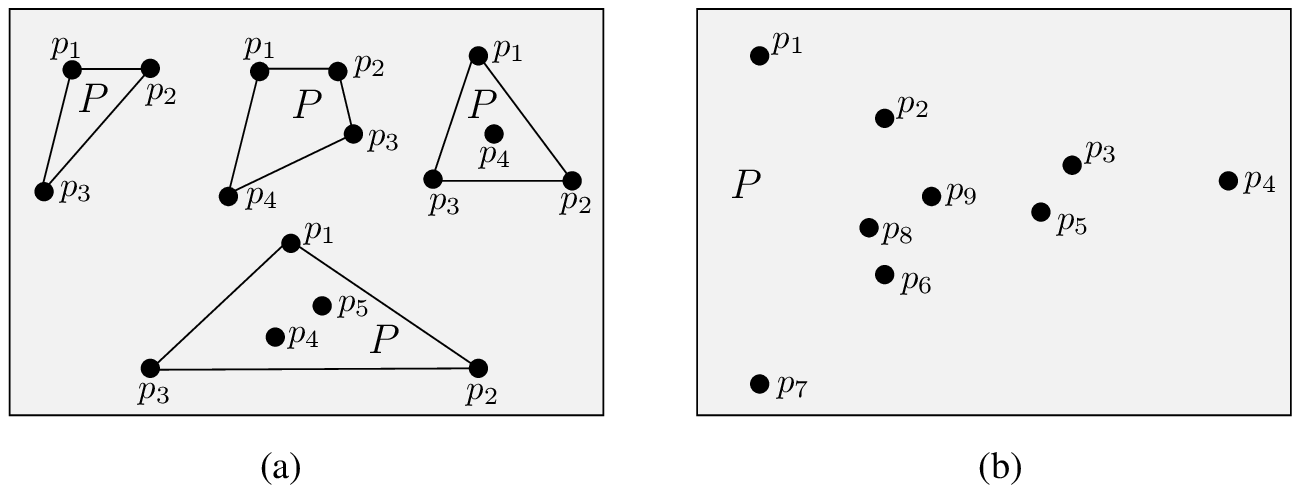}

Observe that  some points of $P-X$ may lie inside $C$, i.e., $C$ may not be empty (see \Fig{lconvexk}(a)).
In this context, Erd\"{o}s \cite{e-smp-1978,e-sagt-1981} posed a
problem of determining the smallest positive integer $h(k)$ (if it exists) such that any point
set $P$ of at least $h(k)$ points in general position in the plane has
$k$ points that are vertices of an empty convex polygon $C$. 
For an empty triangle, $h(3)=3$ and for an empty quadrilateral, 
it can be seen that $h(4)=5$ (\Fig{lconvexk1}(a)). For an
empty pentagon,  \Fig{lconvexk1}(b) demonstrates $h(5) \geq 10$. In fact,
Harborth \cite{h-kfep-1978} proved that $h(5)=10$. For an empty hexagon,
Gerken \cite{g-echp-2008} and Nicol\'{a}s \cite{n-eht-2007}
showed independently that $h(6) \leq g(9) \leq 1717$ and 
$h(6) \leq g(25)$ respectively. On the other hand, 
Overmars \cite{o-fspe-2003}
established a lower bound based on computer experiment that $h(6) \geq 30$. The gap  between the 
bounds has been
reduced by Koshelev \cite{k-espc-2007} by showing that $h(6) \leq max\{g(8),400\} \leq 463$.
For $k \geq 7$, $h(k)$ is not
bounded as shown by Horton \cite{h-swe-1983}.

Let us consider the other situation where point sets $P$ contain collinear points. 
So, the boundary of a convex polygon $C$ formed by
a subset of points $X$ of $P$ may contain collinear points. Let $Y$ be the points
of $X$ that are on corners of $C$. The points of $Y$ are called
{\it points in  strictly
convex position} as deletion of any point of $Y$  reduces the area of $C$. 
The  Erd\"{o}s-Szekeres theorem mentioned above generalises to the following 
theorem \cite{abb-elps-2009,m-ldg-2002}.

\begin{theorem}\label{lconvexstrict}

For every integers  $\ell \geq 2$ and $k \geq 3$, there exists a smallest integer 
$g(k,\ell)$ such that any
point set $P$ of at least $g(k,\ell)$ points in the plane contains\\
(i) $\ell$ collinear points, or\\ (ii) $k$ points in strictly convex position. 

\end{theorem} 

A straightforward  upper bound on  $g(k,\ell)$ can be derived as given in \cite{abb-elps-2009}.
Assume that $P$ has $\ell-1$ collinear points and at most $k-1$ points in strictly 
convex position.
Let $X \subseteq P$ be any maximal set of points in strictly convex position. So, every point
of $P-X$ is collinear with two points in $X$. So, $|X| \choose 2$ lines cover all
points of $P$ and each line can have at most  $\ell-3$ points of $P-X$. Therefore,
$|P| \leq {|X| \choose 2}(\ell-3)+|X| \leq {k \choose 2}(\ell-3)+k-1$.
If one more point is added to $P$ (i.e., $|P| \leq {k \choose 2}(\ell-3)+k)$), 
then $P$ must contain $\ell$ collinear points or  $k$ 
points in strictly convex position. A tighter upper bound on  $g(k,\ell)$ has been derived by 
Abel et al. \cite{abb-elps-2009}.

Observe that $P$ with $g(k,\ell)$ points may have $k$ points in strictly convex position
but the convex polygon $C$ formed by these $k$ points may not be empty. So,  
the visibility graph $G$ of $P$ having $g(k,\ell)$ points 
may not have a clique of size $k$ as some points of $P$ lying inside $C$ may block 
the visibility between vertices of $C$. In the following, we state the 
{\it Big-Line-Big-Clique Conjecture} of
K\'{a}ra et al. \cite{kpw-ocnv-2005}.

\begin{conjecture}\label{lkara2}

For all integers  $k \geq 2$ and $\ell \geq 2$, there is an integer $h(k,\ell)$ such that any
point set $P$ of at least $h(k,\ell)$ points in the plane contains
$\ell$ collinear points, or $k$ mutually visible points. 

\end{conjecture}

It has been shown that a natural approach to settle this conjecture using extremal
graph theory fails \cite{pw-ovb-2010}. On the other hand, the conjecture is 
%true for $k \leq 5$ or $\ell \leq 3$ \cite{kpw-ocnv-2005,abb-elps-2009}. It is 
trivially true for
$\ell \leq 3$ and for all $k$ on any point set $P$ having $k$ points. Based on planar graphs
shown in \Fig{lplanar},
K\'{a}ra et al. \cite{kpw-ocnv-2005} showed that every point set $P$ of at least
$max\{7, \ell +2\}$ points contains $\ell$  collinear points or $4$ mutually visible
points. Using Theorem \ref{lconvexstrict},
Abel et al. \cite{abb-elps-2009} proved that the conjecture is true for $k=5$ and for
all $\ell$. 
%Since the chromatic number of any graph containing a clique of size $k$ is at least $k$,
For weaker versions of  Conjecture \ref{lkara2} 
relating chromatic number with clique size,  see P\'{o}r and Wood \cite{pw-ovb-2010}.

\begin{unsolved}\label{lkara3}
Prove Conjecture \ref{lkara2} for $k=6$ or $\ell=4$. 
\end{unsolved}

\subsection{Blockers of Visibility Graphs}

\AFigure{lblocker1}
{(a) $Q=(q_1, q_2, \ldots, q_7)$ is a blocking set for $P=(p_1, p_2, \ldots, p_5)$.
(b) Four collinear points need three blockers. 
%}{figs/fblocker1.eps}
}{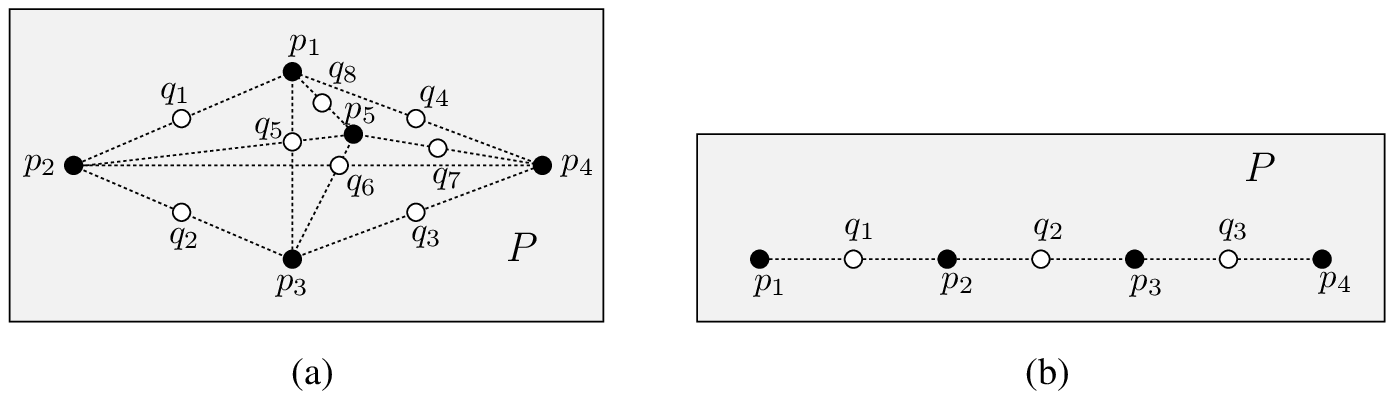}

Let $P$ be a set of $n$ points in the plane. Let $Q=(q_1, q_2, \ldots, q_j)$ be another 
set of points (or blockers) 
in the plane such that (i) $P \cap Q= \emptyset$ and (ii) every segment with both endpoints in $P$
contains at least one point of $Q$ (see \Fig{lblocker1}(a)). In other words, 
there is no edge in the visibility
graph of $P \cup Q$ that connects two points of $P$.  Any such set $Q$ is called a {\it blocking 
set} for $P$. If all points of $P$ are collinear (see \Fig{lblocker1}(b)), then one blocker $q_i$
is placed on the
midpoint of each visible pairs in $P$, and therefore, $|P|-1$ blockers are necessary and
sufficient. On the other hand, what is the minimum size of blocking set $Q$ for $P$ having
no three points collinear (see \Fig{lblocker1}(a))? Note that a blocker may block 
several pairs of visible points
if it is placed on intersection points of segments connecting points of $P$.

Let $b(P)$ denote the smallest size blocking set $Q$ for $P$. Let  $b(n)$ denote the minimum
of $b(P)$ for all $P$ having $n$ points with no three points being collinear. 
It is obvious that $b(n) \geq n-1$. A better lower bound can be derived using a 
triangulation of $P$ \cite{m-bvpg-2009}. Observe that every edge of a triangulation must contain one
blocker. Since every triangulation has at least $2n-3$ edges, it follows that
$b(n) \geq 2n-3$. The lower bound is improved to $b(n) \geq (\frac{25}{8}-o(1))n$ by Dumitrescu
et al. \cite{dpt-nbvp-2009}. 

Let us discuss upper bounds on $b(n)$. It is obvious that  $b(n) \leq {n \choose 2}$. Let
$\mu(P)$ denote the size of the set of midpoints of all ${n \choose 2}$ segments between points
of $P$. Let $\mu(n)$ denote the minimum of $\mu(P)$ for all $P$ having $n$ points with no 
three points being collinear. Using Freiman's theorem on set addition \cite{f-fsts-1973}, 
Stanchescu \cite{ys-psc-2002}
and Pach \cite{p-msip-2003}  have independently
shown that $b(n) \leq \mu(n) \leq n2^{c \sqrt{\log n}}$,
where $c$ is an absolute constant. This shows that if $ \mu(n)$ is not $O(n)$, it can only be slightly
super-linear. 
Many authors \cite{dpt-nbvp-2009,hhuz-mpp-2001, m-bvpg-2009,pw-ovb-2010,ys-psc-2002} have stated or conjectured that
 every set of points $P$
in general position requires a super-linear number of blockers as given below.

\begin{unsolved}\label{lblock1}
Prove that as $n \rightarrow \infty$, $\frac{b(n)}{n} \rightarrow \infty$. 
\end{unsolved}

\AFigure{lkset1}
{Since four colours are required for colouring points in $P$, it is a $4$-blocked point set. 
%}{figs/fkset1.eps}
}{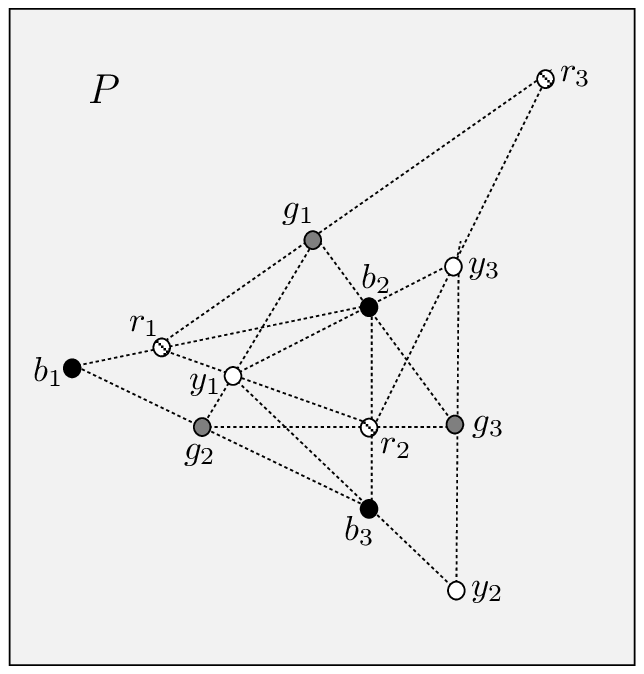}

Let us consider another problem of blockers in a visibility graph introduced by 
Aloupis et al. \cite{abc-bcp-2010}. Let $P$ be a set of points
in the plane with some collinear points (see \Fig{lkset1}). 
The problem is to assign $k \geq 2 $ colours to points 
of $P$ such that (i) if two points are mutually visible in $P$, assign different colours to them,
and (ii) if two points are not visible due to some collinear points in $P$, assign the same 
colour to both of them. Note that this method of colouring is different from the standard 
method of colouring of a graph due to the additional condition (ii). 
Any set of points that admits such a colouring with $k$ colours (for a fixed $k$) is called
a {\it k-blocked point set}. Aloupis et al. \cite{abc-bcp-2010} have made the following conjecture.

\begin{conjecture}\label{lblockc1}
For each integer $k$, there is an integer $n$ such that every $k$-blocked point 
set has at most $n$ points.
\end{conjecture}

Let $p_i$, $p_j$ and $p_l$ be three points in $P$ such that $p_i$ is not visible from both 
$p_j$ and $p_l$. By condition (ii), all three points should be assigned the same colour. 
If $p_j$ and $p_l$ are also not mutually visible due to a collinear point in $P$, then the same 
colour can certainly be assigned to all three of them. However, if  $p_j$ and $p_l$ are 
mutually visible, both conditions (i) and (ii) cannot be satisfied, and therefore, such a 
colouring is not possible which means that $P$ is not a $k$-blocked point set. 
This implies that  points of $P$ that have received 
the same colour must be an independent set in the visibility graph of $P$. In other words,
colour classes correspond to partition of points of $P$ into independent sets. Following lemmas
of  Aloupis et al. \cite{abc-bcp-2010} follow from the above discussion.

\begin{lemma}\label{lcoloured1}

At most three points are collinear in  every $k$-blocked point set. 

\end{lemma}

\begin{lemma}\label{lcoloured2}

Each colour class in a $k$-blocked point set is in a general position. 

\end{lemma}

\AFigure{lkplanar}
{Every $2$-blocked point set has at most $3$ points, and every $3$-blocked point set 
has at most $6$ points
% }{figs/fkplanar.eps}
}{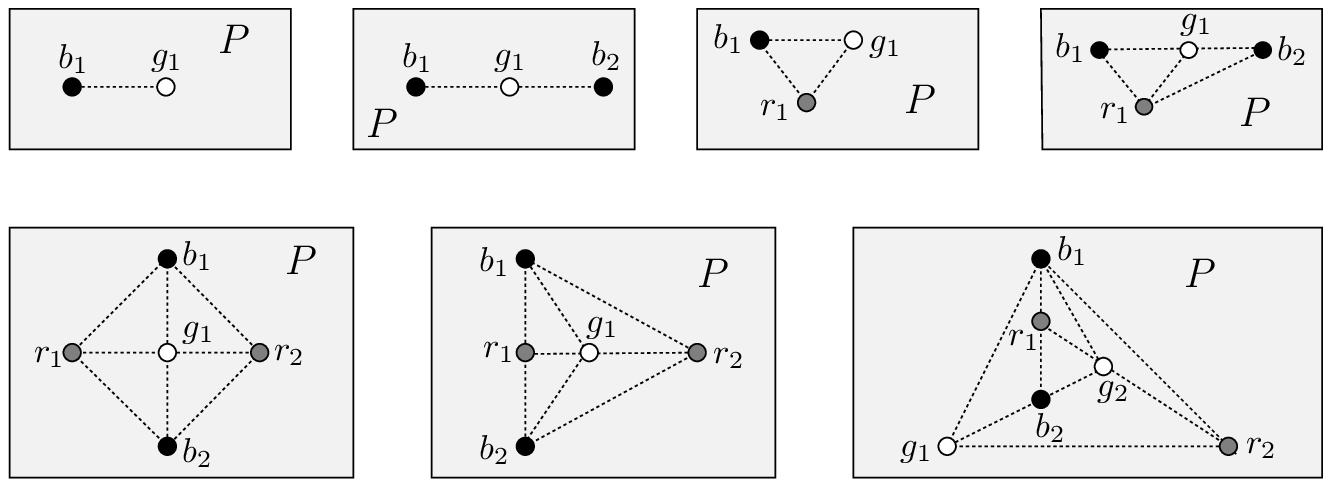}

Let us discuss Conjecture \ref{lblockc1} for different values of $k$. It can be seen
from \Fig{lkplanar} that every $2$-blocked point set has at most $3$ points \cite{abc-bcp-2010}.
It follows from the characterization of $3$-colourable visibility graphs by 
K\'{a}ra et al. \cite{kpw-ocnv-2005} that 
every $3$-blocked point set has at most $6$ points (see \Fig{lkplanar}). 
 Aloupis et al. \cite{abc-bcp-2010} have proved that every $4$-blocked point set has at most 
$12$ points (see \Fig{lkset1}). They also   
%Aloupis et al. \cite{abc-bcp-2010} have 
made the following conjectures on the size of blocked
point sets.

\begin{conjecture}\label{lblockc2}
Every $k$-blocked point set has $O(k^2)$ points. 
\end{conjecture}

\begin{conjecture}\label{lblockc3}
In every $k$-blocked point set, there are at most $k$ points in each colour class.
\end{conjecture}

%\bibliographystyle{plain}
%\bibliography{vis}

%\end{document}

\subsection{Obstacle Representations of Visibility Graphs}

Let $P=(p_1, p_2, \ldots, p_n)$ be a set of points in the plane. Let $Q=(Q_1, Q_2, \ldots, Q_h)$
be a set of simple polygons in the plane called {\it obstacles}.
Construct the visibility graph $G$ such that every point $p_i$ of $P$ is represented as a 
vertex $v_i$ of $G$, and two vertices $v_i$ and $v_j$ of $G$ are connected by an edge in $G$ 
if and only if the line segment $p_ip_j$ does not intersect any obstacle $Q_j$ for all $j$. 
We assume that the line segment joining any two points of $P$ does not pass through a point 
of $P$ or any vertex of an obstacle, i.e., all points of $P$ and vertices of all obstacles are 
in general position. 
We call the pair $(P,Q)$ as obstacle representation of $G$.
Polygonal obstacles can be viewed as a generalization of blockers
of visibility graphs discussed earlier.

Consider the problem of obstacle representation of a given graph $G$ of $n$ vertices, which was
introduced by Alpet et al. \cite{akl-ong-2010}. Draw
every vertex $v_i$  of $G$ as a point $p_i$ in the plane and draw obstacles in such a way 
that every segment $p_ip_j$
intersects an obstacle if and only if $(v_i,v_j)$ is not an edge in $G$.
The {\it obstacle number} of $G$ is the minimum number of obstacles required in any 
obstacle representation of $G$.
Since an obstacle can be placed to block the visibility between each pair of points, 
$n \choose 2$ is an upper bound on the obstacle number of $G$.
We have the following question from Alpet et al. \cite{akl-ong-2010}.

\begin{unsolved}
Is the obstacle number of a graph with $n$ vertices bounded above by a linear function of $n$?
\end{unsolved}

Regarding the lower bound on obstacle numbers,
Alpet et al. \cite{akl-ong-2010} have showed that there exists a graph of $n$ vertices with 
obstacle number  $O(\sqrt{\log n})$, which has been improved
to $O(n/\log ^2 n)$ by Mukkamala et al. \cite{mps-glon-2010}. The bound becomes $O(n/\log  n)$
if the obstacles are restricted to convex polygons. 

\begin{unsolved}
Improve the present lower bound $O(n/\log ^2 n)$ of the obstacle number of a graph with $n$ vertices.
\end{unsolved}

Regarding the graphs with low obstacle numbers,
Alpet et al. \cite{akl-ong-2010}
and Mukkamala et al. \cite{mps-glon-2010} have studied graphs with obstacle numbers
$1$ and $2$.  Mukkamala et al. \cite{mps-glon-2010} showed that for any positive integer $h \geq 3$,
there is a graph with obstacle number exactly $h$. The following questions of 
Alpet et al. \cite{akl-ong-2010} are still open.

\begin{unsolved}
 For $h>1$, what is the smallest number of vertices of a graph with obstacle number $h$?
\end{unsolved}

\begin{unsolved}
Does every planar graph have obstacle number $1$?
\end{unsolved}

\subsection{Connectivity of Visibility Graphs}

Let us consider the problems of vertex-connectivity and edge-connectivity
of visibility graphs. A graph $H$ is called {\it k-vertex-connected} 
(or, {\it k-edge-connected}) if $H$
remains connected even after deleting $k-1$ vertices (respectively, edges)
from $H$. For such a graph $H$, there exists $k$ number of
vertex (or, edge) disjoint paths
between every pair of vertices of $H$ by Menger's theorem \cite{d-gt-2010}.
Let $\kappa(H)$ and $\lambda(H)$ denote the vertex and edge connectivity of $H$ respectively. Let $\delta(H)$ denote the minimum degree of $H$. 
We know that $\kappa(H) \leq \lambda(H) \leq \delta(H)$. Like any graph,
all these properties naturally hold for a visibility graph $G$. 
Here, we state additional
properties on connectivity of visibility graphs given by Payne et al. \cite{2011arXiv1106.3622P}.

We know that the distance between two vertices in a graph is 
the number of edges in the shortest path between them, and the diameter
of the graph is the longest path among the shortest paths between
every pair of vertices in the graph. If all points of $P$ are not collinear,
then every pair points in $P$ must be visible from some  point of $P$.
So, the shortest path between any two non-adjacent vertices in a 
visibility graph $G$ passes through exactly two edges of $G$. Therefore, 
the diameter of  $G$ is of length $2$ \cite{kpw-ocnv-2005}. On the other
hand, Plesn\'{i}k \cite{p-cggd-1975}
has proved that the edge-connectivity of a graph
with the diameter of length at most $2$ equals its minimum degree. So, we have 
$\kappa(G) \leq \lambda(G)=\delta(G)$. This result has been strengthen by
Payne et al. \cite{2011arXiv1106.3622P} for visibility graphs as stated 
in the following theorems.

\begin{theorem}

Let $v_i$ and $v_j$ be two vertices of a visibility graph $G$. If
$d$ is the minimum of the degree of $v_i$ and the degree of $v_j$,
then there exists $d$ number of edge disjoint paths of length at most 4
between $v_i$ and $v_j$ in G.

\end{theorem}

%\begin{theorem}
%Every minimum edge-cut in a visibility graph $G$ is the set of edges incident
%to the same vertex of $G$.
%\end{theorem}

\begin{theorem}
Let $S$ be a set of minimum number of edges in a visibility graph $G$
whose removal disconnects $G$. Then, all edges of $S$ are incident to the 
same vertex of $G$.
\end{theorem}

\begin{theorem}
Every visibility graph $G$ with minimum degree $\delta(G)$ has vertex-connectivity 
at least $\frac{\delta(G)}{2}+1$.
\end{theorem}

\begin{theorem}
Every visibility graph $G$ with minimum degree $\delta(G)$ has 
vertex-connectivity 
at least $\frac{2 \delta(G) +1}{3}$ if 
the number of collinear points of $P$ on a line is restricted to $4$.
\end{theorem}

\AFigure{lconnectivity}
{The black vertices are cut set for a visibility graph with vertex-connectivity
$\frac{2 \delta(G) +1}{3}$. The minimum degree  $\delta(G)=3k+1$ is achived
for extreme corner points. Not all edges are drawn in the figure.
%}{figs/fconnectivity.eps}
}{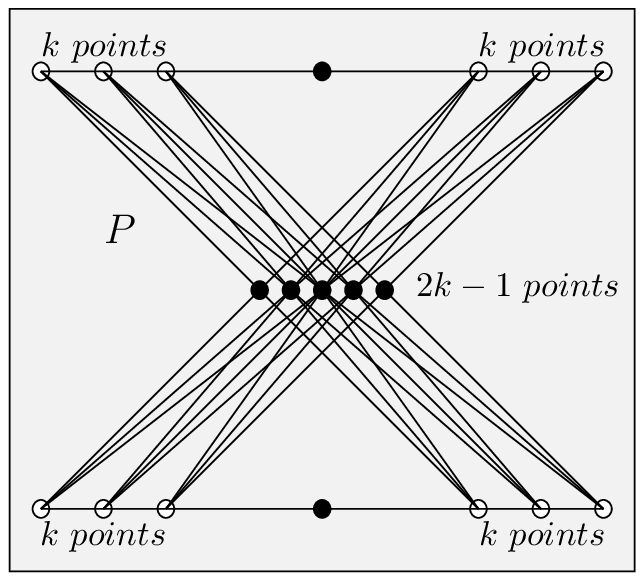}

\begin{conjecture}
Every visibility graph $G$ with minimum degree $\delta(G)$ has vertex-connectivity 
at least $\frac{2 \delta(G) +1}{3}$ (see \Fig{lconnectivity}).
\end{conjecture}

\section{Visibility graph theory: Segments}

%\subsection{Visibility Graph Recognition, Characterization, and Reconstruction}
\subsection{Visibility Graphs: Recognition, Characterization, and Reconstruction}

In Section \ref{lintroduction}, we have defined 
the segment visibility graph $G$ for a given set of disjoint line segments 
$S$ (see \Fig{lvisgraph-segment}(b)).
We have also stated how to compute $G$ from $S$ efficiently. 
Consider the opposite problem of determining if there is a set of 
disjoint segment $S$ 
whose visibility graph is the given graph $G$. 
This problem is called the segment visibility graph 
{\it recognition} problem.  Identifying the set of properties satisfied by all segment visibility 
graphs is called the segment visibility graph {\it characterization} problem. The problem of 
actually drawing one such set of segment $S$ whose visibility graph is the given graph $G$, 
is called the segment visibility graph {\it reconstruction} problem. 

All three above problems are open for segment visibility graphs.  
Only characterization known is for a sub-class  
given by Everette et al. \cite{ehkn-cgta2-2000}.
They have characterized those segment visibility graphs that do not have $K_5$ (a complete graph of
five vertices) as a minor. A graph $M$ is called a {\it minor} of a graph $G$ if $M$ can be obtained
from $G$ by a sequence of vertex deletions, edge deletions and edge contractions. Their characterization
gives a straightforward polynomial time algorithm for recognizing this class of graphs.   

\begin{unsolved}
 Given a graph $G$ in adjacency matrix form, determine whether $G$ is the segment visibility graph 
of a set of disjoint segments $S$ in the plane. 
\end{unsolved}

\begin{unsolved}
Characterize the segment visibility graphs.
\end{unsolved}

\begin{unsolved}
Given a segment visibility graph $G$, draw the segments $S$ in the plane 
whose visibility graph is $G$.
\end{unsolved}

\subsection{Hamiltonian Cycles in Visibility Graphs}

Let $G$ be the segment visibility graph of a given set $S$ of disjoint line segments.
Consider the problem of identifying a Hamiltonian cycle $C$ in $G$
(see  \Fig{lhamilton1-segment}). 
A cycle in $G$ is called a {\it Hamiltonian cycle}
if the cycle passes through all vertices of $G$ exactly once. There can be two types of
Hamiltonian cycles $C$ in $G$. Assume that $G$ is drawn directly on the segments of $S$
and call this embedded segment visibility graph as $G'$.
If no two segments in $G'$ corresponding to edges of a cycle $C$  intersect,
then $C$ forms the boundary of a simple polygon in $G'$. Such cycles are called
{\it Hamiltonian circuits} \cite{r-cscfs-89} (see \Fig{lhamilton1-segment}(b)), 
and the corresponding polygons are called
{\it spanning polygons} \cite{m-htcp-1992} or {\it Hamiltonian polygons} \cite{ht-sevg-2003}.
Otherwise, $C$ corresponds to the boundary of a self-crossing polygon in  $G'$
(see \Fig{lhamilton1-segment}(a)).
A  Hamiltonian polygon $Q$ is called a {\it circumscribing polygon} if it has an additional
property that no segment of $S$ lies to the exterior of $Q$, i.e., each segment of $S$
is either an edge on the boundary of $Q$ or an internal chord of $Q$ \cite{m-htcp-1992}
(see  \Fig{lhamilton1-segment}(b)).   
Mirzaian \cite{m-htcp-1992} made the following conjectures.

\AFigure{lhamilton1-segment}{
  (a) The cycle $C=(p_1, p_{10}, p_9, p_8, p_7, p_5, p_4, p_6, p_3, p_2, p_1)$ is Hamiltonian 
but is self-intersecting.
  (b) The cycle $C=(p_1, p_9, p_{10}, p_7, p_8, p_6, p_5, p_4, p_3, p_2, p_1)$ is a Hamiltonian 
circuit as it forms a boundary of a simple polygon.
  (c) There is no alternating cycle as both endpoints of the segment $p_9p_{10}$ belong
to the convex hull of the segments \cite{rit-cscf-90}.
 % }{figs/fhamilton1.eps}
  }{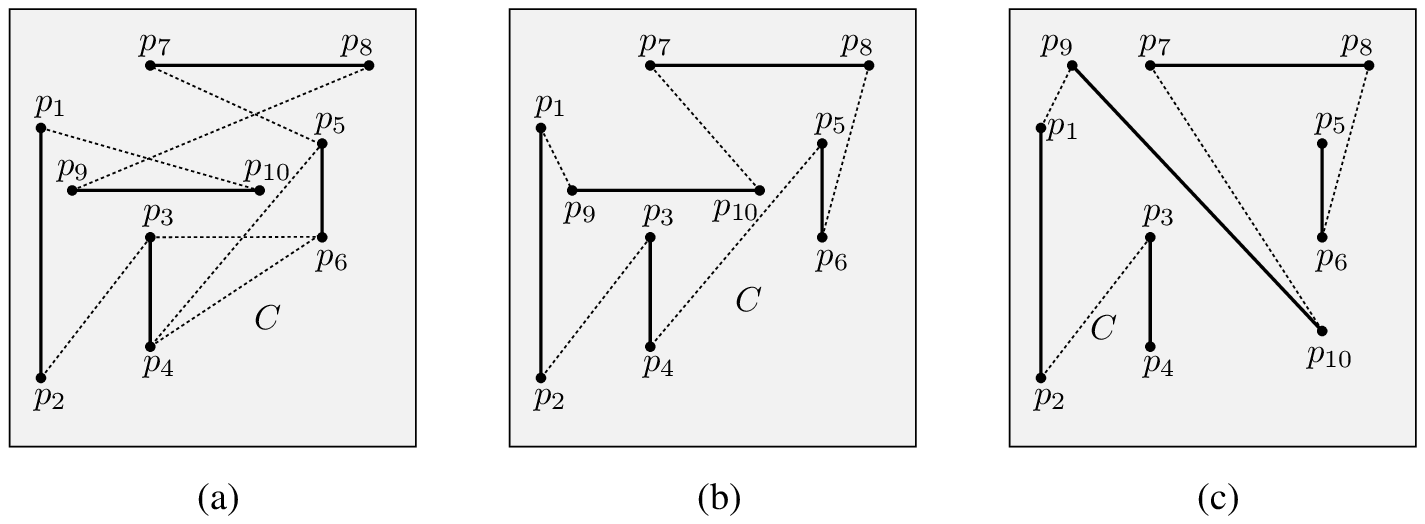}

\begin{conjecture}\label{lspann1}

Every segment visibility graph $G$ contains a Hamiltonian cycle $C$.
 
\end{conjecture}

\begin{conjecture}\label{lspann2}

Every segment visibility graph $G$ contains a Hamiltonian cycle $C$ that corresponds to a
Hamiltonian circuit in the embedded segment visibility graph $G'$.
 
\end{conjecture}

\begin{conjecture}\label{lspann3}

Every segment visibility graph $G$ contains a Hamiltonian cycle $C$ that corresponds to
the boundary of a circumscribing polygon in the embedded segment visibility graph $G'$.
 
\end{conjecture}

Observe that if Conjecture \ref{lspann3} is true, then both Conjectures \ref{lspann2} and \ref{lspann1}
are also true. However,  Urbe and Watanabe \cite{uw-occm-1992} gave a counter-example
to Conjecture \ref{lspann3}. For special classes of
segments in $S$, Conjectures \ref{lspann2} was proved  by 
Mirzaian \cite{m-htcp-1992} and O'Rourke and J. Rippel \cite{or-tschv-94}.
Later, Conjectures \ref{lspann2} was proved for all classes of segments in $S$ 
by Hoffmann and T\'{o}th \cite{ht-sevg-2003} using the result of 
Bose et al. \cite{bht-esdl-2001}, and they presented 
an $O(n \log n)$ time algorithm for locating a Hamiltonian circuit in $G'$.  
Since Conjectures \ref{lspann2} is true, Conjectures  \ref{lspann1} is also true. 

Observe that the algorithm of Hoffmann and T\'{o}th \cite{ht-sevg-2003} takes $S$ as an input
and then locates a Hamiltonian circuit in the embedded segment visibility graph $G'$. 
Suppose $S$ is not given as an input but
only $G$ is given, then it is not clear how to identify a Hamiltonian cycle
in $G$ in polynomial time as there is no known algorithm for segment visibility graph 
reconstruction problem. So, we have the following problems.

\begin{unsolved}
 Given a segment visibility graph $G$ in adjacency matrix form, identify a Hamiltonian
cycle in $G$ in polynomial time.
\end{unsolved}

\begin{unsolved}
 Given a segment visibility graph $G$ in adjacency matrix form, identify the edges of
$G$ that correspond to segments of $S$.

\end{unsolved}

Consider the problem of identifying a special type of Hamiltonian circuit $C$ in $G'$
where every alternate edge of $C$ is a segment of $S$ (see \Fig{lhamilton1-segment}(b)). 
It has been shown by Rappaport
et al. \cite{rit-cscf-90} that such an {\it alternating cycle} may not 
always exist in $G'$ (see \Fig{lhamilton1-segment}(c)). On the other hand, 
they showed that an alternating
cycle always exists if one endpoint of every segment in $S$ belongs to the convex hull
of $S$. For this special class of segments, they gave an $O(n \log n)$ time algorithm for 
constructing an alternating cycle.

If $G'$ does not contain an alternating cycle, it is natural to ask for 
a longest alternating path that is present in $G'$ (see \Fig{lhamilton1-segment}(c)). 
Urratia \cite{u-opcg-2002} made the
following conjecture which was proved by Hoffmann and T\'{o}th \cite{ht-aptd-2003}.

\begin{conjecture}\label{lalter1}
In the embedded segment visibility graph $G'$ of a set  $S$ of $n$ disjoint segments, 
there exists an alternating path containing at least  $O( \log n)$ segments of $S$.

\end{conjecture}

\subsection{Bar Visibility Graphs}

The idea of representing a graph using a visibility relation 
was introduced in the 1980s as a model 
tool for VLSI layout problems \cite{dhvm-rpgv-1983,slmlw-vpvl-1985}.
A graph $G$ is called a {\it bar visibility graph} if its 
vertices $v_1, v_2, \ldots, v_n$ can be associated
with a set $S$ of disjoint line segments (or, horizontal bars) $s_1, s_2, \ldots, s_n$ 
in the plane such 
that $v_i$ and $v_j$
are joined by an edge in $G$ if and only if there exists an unobstructed vertical line of sight
between $s_i$ with $s_j$ \cite{tt-dcg-86}. The set  $S$ is called 
a {\it bar visibility representation} of 
$G$  (see \Fig{lbar-visibility}). 
If each line of sight is required to be a rectangle of positive width, then $S$ is an
$\epsilon$-{\it visibility representation} of $G$ (see \Fig{lbar-visibility}(b)). 
If each line of sight is a segment
(i.e. width is $0$), then $S$ is  a {\it strong visibility representation} 
of $G$ (see \Fig{lbar-visibility}(c)). Note that if a vertical segment between two
horizontal segments (say, $s_1$ and $s_3$ in \Fig{lbar-visibility}(c))
passes through
an endpoint of another horizontal segment (i.e., $s_2$), it is considered 
that the line of sight is obstructed
by the middle horizontal segment. 
%For example, the vertical segment connecting the right
%endpoints of $s_1$ and $s_3$ in \Fig{lbar-visibility}(c) passes through the right endpoint of $s_2$
%and therefore, $s_1$ and $s_3$ are not visibile.
We have the following theorems on the 
characterizations and  representations
of bar visibility graphs \cite{an-dam-92,Kw-dbr-96,lmw-nvg-1987,rt-rplb-1986,tt-dcg-86,ws-socg-1985}.

\AFigure{lbar-visibility}{
  (a) A given graph $G$. 
  (b) An $\epsilon$-visibility representation of $G$.
  (c) A strong visibility representation of $G$.
  %}{figs/fbar-visibility.eps}
   }{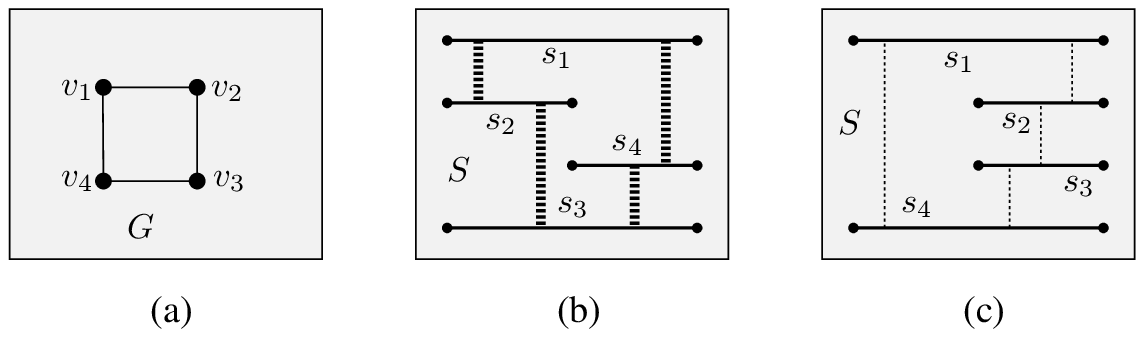}

\begin{theorem}

A graph $G$ admits an $\epsilon$-visibility representation if and only if there is 
a planar embedding of $G$ such that all cutpoints of $G$ appear on the boundary of the 
external face in the embedding.
 
\end{theorem}

\begin{theorem}

An $\epsilon$-visibility representation of a $2$-connected planar graph $G$ of $n$ vertices
 can be done in $O(n)$ time.

\end{theorem}

\begin{theorem}

Let $G$ be a $2$-connected planar graph. If $G$ admits a  strong visibility representation, 
then there is
no pair of non-adjacent vertices $v_i$ and $v_j$  of $G$ such that the removal of $v_i$ and $v_j$
separates $G$ into four or more components.
  
\end{theorem}

\noindent{\it Proof:} Let $c_1, c_2, \ldots, c_k$ for $k \geq 4$ be the connected component
of $G$ after removing $v_i$ and $v_j$ (see \Fig{lbar-proof}(a)). 
Assume on the contrary that $G$ admits a strong visibility
representation (say, $R$). Let $s_m$ denote the horizontal segment in $R$ corresponding to
$v_m$ of $G$. 

Consider the situation where there exists a vertical line $L$ 
between $s_i$ and $s_j$ of $R$ that does not intersect either of them
(see \Fig{lbar-proof}(b)). Since every component
$c_l$ connects $v_i$ to $v_j$ in $G$, there exists a vertex $v_m$ of $c_l$ for all $l$
such that $L$  intersects the horizontal segment $s_m$ of $R$. This intersection of $L$ with
$s_m$ for every component implies that vertices of different components are connected
by edges in $G$, which is  a contradiction. 

\AFigure{lbar-proof}{
  (a) A given graph $G$. 
  (b) $L$ does not intersect $s_i$ and $s_j$.
  (c) $L$ intersects both $s_i$ and $s_j$.
  %}{figs/fbar-proof.eps}
   }{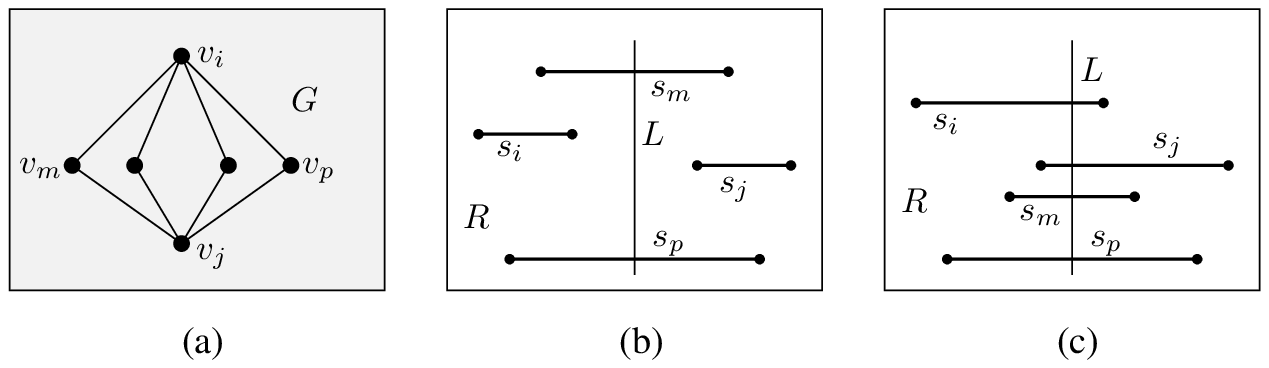}

Consider the other situation where 
there exists a vertical line $L$ intersecting both $s_i$ and $s_j$ of 
$R$ (see \Fig{lbar-proof}(c)). So, there are
three parts of $L$ due to these intersections. If any of these parts of $L$ intersects
two horizontal segments $s_m$ and $s_p$ in $R$ where $v_m$ and $v_p$ belong to different
component in $G$, then there must be  an edge between  $v_m$ and $v_p$  in $G$, which
is a contradiction. So, each part of $L$ can interest horizontal segments of $R$ coming
only from the same component. Therefore, $L$ cannot intersect  horizontal segments 
of $R$ coming from more that three different components. 
Hence, $G$ does not admit a strong visibility representation, 
a contradiction. \hfill{$\Box$}

\begin{theorem}

There exists a $3$-connected planar graph $G$ that does not admit  a strong 
visibility representation.

%All $3$-connected planar graphs are not bar visibility graphs.

\end{theorem}

\begin{theorem}

Every $4$-connected planar graph $G$ admits a strong visibility representation.

\end{theorem}

\begin{theorem}

A strong visibility representation of a $4$-connected planar graph $G$ of $n$ vertices
 can be done in $O(n^3)$ time.

\end{theorem}

Let us consider variations of bar visibility graphs. While representing vertices of $G$ as 
bars, there is no restriction on the length of horizontal bars. Suppose, a restriction
is imposed that all bars in a visibility representation must have the same length.
In that case, we get another type of visibility graphs which are known as
{\it unit bar visibility graphs}.  Though there are characterizations for special classes
of unit bar visibility graphs \cite{chks-cbvt-2006,dgh-unvl-2004,dv-ubvg-2003}, 
no characterization is known for general graphs. 
We have the following problems.

\begin{unsolved}
Characterize unit bar visibility graphs.
\end{unsolved}

\begin{unsolved}
Is recognition of  unit bar visibility graphs NP-complete?
\end{unsolved}

Dean et al. \cite{deglst-bvg-2007} introduced another variation of bar visibility graphs 
called {\it bar k-visibility graphs}, where bars are allowed to see vertically 
through at most $k$ bars under strong visibility (see \Fig{lbar-kvisibility}).
It means that standard 
bar visibility graphs become bar $0$-visibility graphs. If bars are allowed to see vertically 
through all other bars (i.e. bar $\infty$-visibility graphs), then the bar representation gives 
an interval graph representation. Formally, a graph $G$ is called a {\it bar k-visibility graph} 
if its vertices $v_1, v_2, \ldots, v_n$ can be associated with a set $S$ of disjoint 
line segments (or, horizontal bars) $s_1, s_2, \ldots, s_n$ in the plane such 
that $v_i$ and $v_j$ are joined by an edge in $G$ if and only if a vertical segment between 
$s_i$ and $s_j$ intersects at most $k$ segments 
of $S$ \cite{deglst-bvg-2007,fw-tbvg-2007,hvw-frbvg-2007}.

\AFigure{lbar-kvisibility}{
  (a) A given graph $G$. 
  (b) Representation shows that $G$ is a bar $1$-visibility graph.
 % }{figs/fbar-kvisibility.eps}
  }{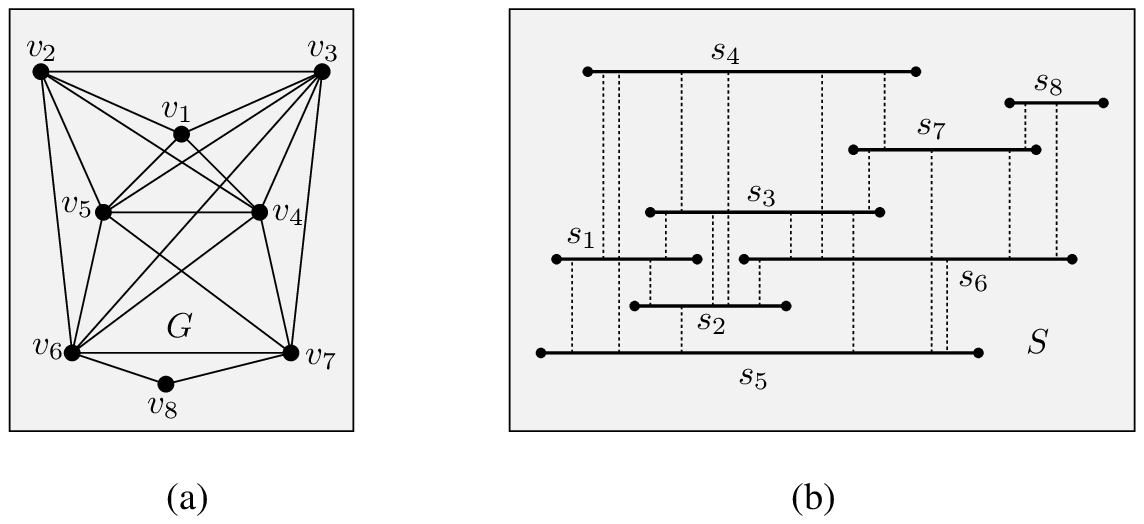}

Unlike bar visibility graphs,
bar $k$-visibility graphs  have not been completely characterized. For understanding this
class of graphs, Dean et al. \cite{deglst-bvg-2007} considered the problem of
bounding the number of edges of a bar $k$-visibility graph $G$. They proved an upper 
bounds of $(k+1)(3n-7/2 k-5)-1$ edges
for $G$ and conjectured an improve edge bound of 
$(k+1)(3n-4 k-6)$ which was proved later by 
Hartke et al. \cite{hvw-frbvg-2007}. They also studied $d$-regular  $k$-visibility graphs.
Dean et al. \cite{deglst-bvg-2007} also proved that $6k+6$ is an upper bound on the chromatic 
number of $G$. We have the following problems \cite{deglst-bvg-2007,hvw-frbvg-2007}.

\begin{unsolved}
Characterize bar $k$-visibility graphs.
\end{unsolved}

\begin{unsolved}
Triangle-free non-planar graphs are forbidden subgraphs of bar $k$-visibility graphs.
Is there any other class of forbidden subgraphs of bar $k$-visibility graphs?
\end{unsolved}

\begin{unsolved}
Are there $(2k+2)$-regular bar $k$-visibility graphs for $k \geq 5$?
\end{unsolved}

\begin{unsolved}
Are there $d$-regular bar $k$-visibility graphs with $d \geq 2k+3$?
\end{unsolved}

\begin{unsolved}
Improve the current upper bound of $6k+6$ on the chromatic number of a bar $k$-visibility graph.
\end{unsolved}

Dean et al. \cite{deglst-bvg-2007} also studied the thickness of bar $k$-visibility graphs.
The {\it thickness} of a graph $G$ is defined as the minimum number of planar graphs whose
union is $G$.  Exact values of thickness is know for very few classes of graphs \cite{mos-tgs-1998}.
Dean et al. proved  an upper bound of $4$  for the thickness of 
bar $1$-visibility graphs, and conjectured that bar $1$-visibility graphs actually have
thickness of at most $2$. The conjecture  was disproved by Felsner and Massow \cite{fw-tbvg-2007}
by constructing a bar $1$-visibility graph having thickness $3$. For a special
class of bar $1$-visibility graphs, Felsner and Massow \cite{fw-tbvg-2007} presented
an algorithm for partitioning the edges into two planar graphs showing that the thickness
of this special class of graphs is $2$.
We have the following problems \cite{deglst-bvg-2007}.

\begin{unsolved}
It has been shown that the thickness of
 a bar $k$-visibility graph is bounded by $2k(9k-1)$. Can this upper bound be improved?
\end{unsolved}

\begin{unsolved}
The crossing number of a graph  is the minimum possible number of crossings with which the graph 
can be drawn in the plane.
What is the largest crossing number of a bar $k$-visibility graph?
\end{unsolved}

\begin{unsolved}
The genus of a graph is the minimal integer $g$ such that the graph can be embedded on a surface of 
genus $g$.
What is the largest genus of a bar $k$-visibility graph?
\end{unsolved}

Bar visibility graphs have been generalized to rectangle visibility 
graphs by considering both vertical and horizonal visibility among bars 
having non-zero 
thickness \cite{bdhs-orvg-1997,dh-rvrbp-1997,o-opcv-1998,r-drvg-1997,sw-rvgc-2003}.
A graph $G$ is called a {\it rectangle visibility graph} if it can be realized by closed isothetic
rectangles in the plane, with pairwise disjoint interiors, with vertices representing rectangles
in such a way that two vertices $v_i$ and $v_j$ of $G$ are connected by an edge if and only if
their corresponding rectangles are vertically or horizontally visible 
from each other by a beam of unobstructed visibility of finite width. 
Unlike bar visibility graphs, no characterization of rectangle visibility graphs are known,
and moreover, Shermer \cite{s-orvsiii-1996} has shown that the problem of recognizing 
them is NP-complete. We have the following problem.

\begin{unsolved}
Characterize rectangle visibility graphs.
\end{unsolved}

\section{Visibility graph theory: Polygons}

%\subsection{Visibility Graphs: Recognition, Characterization, and Reconstruction}
%\subsection{Visibility Graph Recognition, Characterization, and Reconstruction}
%Consider the opposite problem of determining if there is a polygon $P$ whose visibility graph is 
%the given graph $G$. This problem is called the visibility graph {\it recognition} problem. 
%Identifying the set of properties satisfied by all visibility graphs is called the visibility graph 
%{\it characterization} problem. The problem of actually drawing one such polygon $P$ whose 
%visibility graph is the given graph $G$, is called the visibility graph {\it reconstruction} problem. 

%\subsubsection{Visibility Graph Recognition}
\subsection{Visibility Graph Recognition}
In Section \ref{lintroduction}, we have defined 
the vertex visibility graph $G$ for a given simple polygon 
$P$ (see \Fig{lvisgraph}(b)).
We have also stated how to compute $G$ from $P$ efficiently. 
Consider the opposite problem of determining if there is a simple polygon $P$ 
whose visibility graph is the given graph $G$. 
This problem is called the  visibility graph 
{\it recognition} problem for polygons.
The general problem of recognizing a given graph $G$ as the visibility graph of a simple polygon $P$ is
yet to be solved. However, this problem has been solved for visibility graphs of \emph{spiral} polygons
\cite{ev-vgc-90,ec-rvgsp-90} and \emph{tower} polygons \cite{ChoiSC95}.

\begin{unsolved}
 Given a graph $G$ in adjacency matrix form, determine whether $G$ is the visibility graph of a simple 
polygon $P$.
\end{unsolved}

\begin{unsolved}
 Is the problem of recognizing visibility graphs in $NP$?
\end{unsolved}

Ghosh \cite{g-rcvr-86,g-vap-07} presented three necessary conditions for recognizing visibility 
graphs $G$ of a simple polygon $P$ under the assumption that a Hamiltonian cycle $C$ of 
$G$, which corresponds to the boundary of $P$, is given as input along with $G$. 
It can be seen that this problem is easier than the actual
recognition problem as the edges of $G$  corresponding to boundary edges of $P$
have already been identified. Assume that the vertices of $G$ are
labeled with $v_1, v_2, \ldots, v_n$ and 
$C=(v_1, v_2, \ldots, v_n)$ is  in counterclockwise order.
%A {\it cycle} is a simple and closed path in $G$. 
An edge in $G$ connecting two non-adjacent
vertices of a cycle is called a {\it chord} of the cycle.
A cycle $w_1, w_2, \ldots, w_k$ in $G$ is
called {\it ordered} if the vertices
$w_1, w_2, \ldots, w_k$ follow the order in $C$.
The Hamiltonian cycle $C$ is an ordered cycle  of all $n$ vertices in $G$.

\noindent{\bf Necessary condition 1.}
\emph{Every ordered cycle of $k \geq 4$ vertices in a visibility graph $G$ of a simple polygon $P$
has at least $k-3$ chords.}

\noindent{\it Proof:} Since an ordered cycle of $k$ vertices in $G$ 
 corresponds to a sub-polygon $P'$ of $k$ vertices in $P$, 
the ordered cycle must have at least 
$k-3$ chords in $G$ as $P'$ needs $k-3$ diagonals 
for triangulation of $P'$. \hfill{$\Box$}

\AFigure{lcon2}{
   (a) For the invisible pair $(v_4,v_6)$, $v_5$ is the blocking vertex.
   (b) The invisible pair $(v_2,v_5)$ does not have any blocking vertex.
   (c) The blocking vertex  $v_3$ cannot simultaneously block the visibility of
       invisible pairs $(v_2,v_8)$ and $(v_4,v_7)$.
%}{figs/fcon2.eps}
}{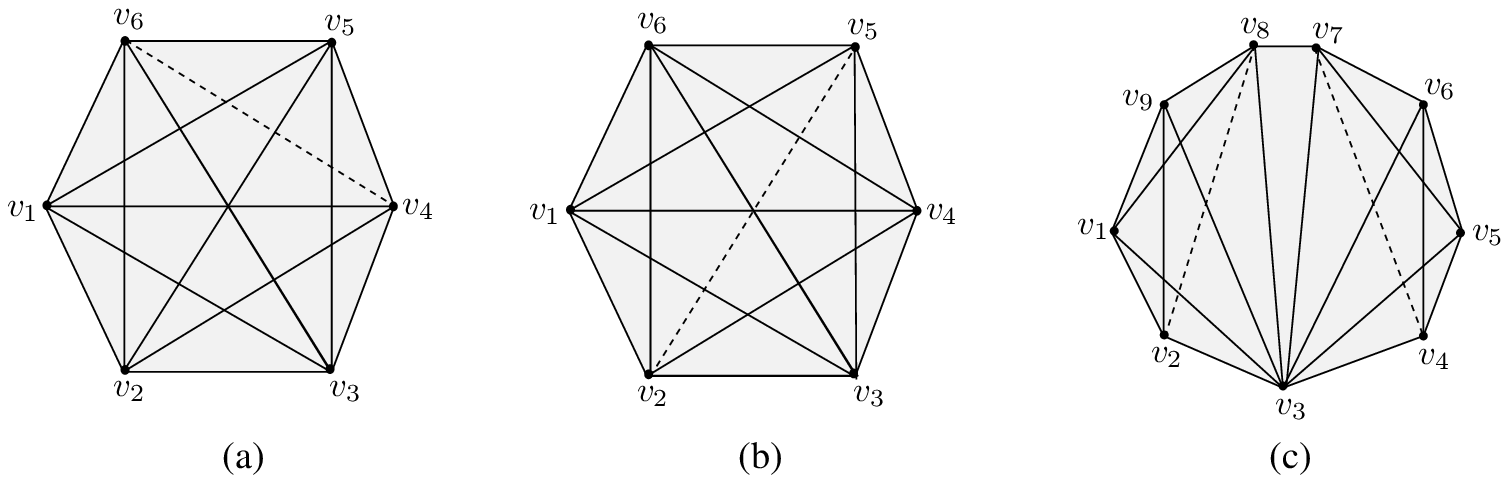}

A pair of vertices $(v_i,v_j)$ in $G$ is a {\it visible} 
 {\it pair} (or {\it invisible pair}) if $v_i$ and $v_j$ are adjacent
(respectively, not adjacent) in $G$. The vertices from $v_i$ to 
$v_j$ on $C$ in counterclockwise order are denoted as $chain(v_i,v_j)$.
Let $v_a$ be a vertex of $chain(v_i,v_j)$ for $i < j$ such that no two vertices 
$v_k \in chain (v_i,v_{a-1})$ and $v_m \in chain (v_{a+1},v_j)$ are connected by an edge 
in $G$. Then $v_a$ is called a {\it blocking vertex} for the invisible 
pair $(v_i,v_j)$ (see \Fig{lcon2}(a) and \Fig{lcon2}(b)).
Intuitively, blocking vertices 
correspond to reflex vertices of the polygon though
all blocking vertices in $G$
may not be reflex vertices in $P$.

\noindent {\bf Necessary condition 2.}
\emph{Every invisible pair $(v_i,v_j)$ in the visibility graph $G$ of a simple polygon $P$
has at least one blocking vertex.}

\noindent{\it Proof:} Since $(v_i,v_j)$ is an invisible pair in $G$,
the {\it Euclidean
shortest path} 
in $P$ between $v_i$ and $v_j$ makes turns
at reflex vertices of $P$, and therefore, 
each of these reflex vertices is a blocking
vertex for $(v_i,v_j)$ in $G$.
\hfill{$\Box$}

Let $v_a$ be a blocking vertex in $G$ for
two invisible pairs $(v_i,v_j)$ and $(v_k,v_l)$.
Traverse the Hamiltonian cycle $C$  from $v_a$
in counterclockwise order. If both $v_k$ and $v_l$ 
are encountered before $v_i$ and $v_j$ during the traversal, then 
$(v_i,v_j)$ and $(v_k,v_l)$ are referred as {\it separable} with respect to $v_a$.
In \Fig{lcon2}(c), invisible pairs $(v_2,v_8)$ and $(v_4,v_7)$ are separable with
       respect to the blocking vertex $v_3$.

\noindent{\bf Necessary condition 3.} 
{\it Two separable invisible pairs $(v_i,v_j)$ and $(v_k,v_l)$ in the visibility graph 
$G$ of a simple polygon $P$ must have distinct 
blocking vertices.}

\noindent{\it Proof:} Let $(v_i,v_j)$ and $(v_k,v_l)$ be two separable invisible pairs and
the vertex $v_a$ is their sole
blocking vertex. So, $v_a$ must be a reflex vertex in $P$. 
Since the visibility in $P$ between $v_i$ and $v_j$
as well as between $v_k$ and $v_l$ can only be blocked by $v_a$ and
the sub-polygons of $P$ corresponding to  ordered cycles 
$v_i, v_a, v_j, \ldots, v_i$ and $v_a, v_k, \ldots, v_l, v_a$ are disjoint, 
$v_a$ cannot simultaneously block the visibility between $v_i$ and $v_j$
and between $v_k$ and $v_l$ in $P$.
\hfill{$\Box$}

It has been pointed out by Everett and Corneil \cite{ev-vgc-90,ec-nrcvg-95} that these three 
conditions are not 
sufficient as there are graphs that satisfy the three necessary conditions but  are not visibility 
graphs of any simple polygon. These counterexamples can be eliminated once the third necessary 
condition is strengthened. It have been shown by Srinivasraghavan and Mukhopadhyay \cite{sm-nncv-94}
that the stronger version of the third necessary condition proposed by Everett \cite{ev-vgc-90} is 
in fact necessary. 

\noindent{\bf Necessary condition 3${\boldmath '}$.}
\emph{There is an assignment in a visibility graph   such that no blocking vertex $v_a$ is assigned to 
two or more minimal invisible pairs that are separable with respect to $v_a$.}

On the other hand, the counterexample given by Abello, Lin and Pisupati 
\cite{akp-vgsp-90} shows that the three necessary conditions of Ghosh \cite{g-rcvr-86} are not 
sufficient even with the stronger version of the third necessary condition. In a later paper by 
Ghosh \cite{g-rcvg-97}, another necessary condition is identified which circumvents the counterexample 
of  Abello, Lin and Pisupati \cite{akp-vgsp-90}. 
%Ghosh's necessary conditions are as follows \cite{g-vap-07}.

\noindent{\bf Necessary condition 4.}
\emph{Let $D$ be any ordered cycle of the visibility graph $G$ of a simple polygon $P$. 
For any assignment of blocking vertices to all 
minimal invisible pairs in $G$, the total number of vertices of $D$ assigned to the 
minimal invisible pairs between the vertices of $D$ is at most $|D|-3$.}

\noindent{\it Proof:} Let $P'$ be the subpolygon of $P$ such that the boundary of $P'$
corresponds to $D$.
If every blocking vertex $v_a \in D$ is assigned to some minimal invisible pair 
between vertices of $D$, 
$v_a$ becomes a reflex vertex in $P'$.
So, the sum of internal angles of $P'$
is more than $(|D|-2)180^\circ$ contradicting the fact that
the sum of internal angles of any simple polygon of $|D|$
vertices is $(|D|-2)180^\circ$.
\hfill{$\Box$}

It has been shown later by Streinu \cite{is-cccg-1999,is-cgta-2003} that these four necessary conditions 
are also not sufficient. It is not clear whether another necessary condition is required to circumvent the 
counter example. For more details on the recognition of visibility graphs, see Ghosh \cite{g-vap-07}.

\begin{unsolved}
  Given a graph $G$ in adjacency matrix form along with a Hamiltonian cycle $C$ of $G$, 
determine whether $G$ is the visibility graph of a simple polygon $P$ whose boundary corresponds 
to the given Hamiltonian cycle $C$.
\end{unsolved}

Everett \cite{ev-vgc-90} presented an $O(n^3)$ time algorithm for testing Necessary Condition 1
which was later improved by Ghosh \cite{g-rcvg-97,g-vap-07} to $O(n^2)$ time. 
Ghosh also gave an $O (n^2)$ time algorithm for testing Necessary Condition 2. Das, Goswami
and Nandy \cite{dgn-tncr-2002} showed that Necessary Condition 3$'$ can be tested in $O(n^4)$ time.
We have the following theorem.
\begin{theorem}
Given a graph $G$ of $n$ vertices and a Hamiltonian cycle $C$ in $G$,
Necessary Conditions 1, 2 and $3'$ can be tested in $O (n^4)$ time. 

\end{theorem}

\begin{unsolved}

Design an algorithm for testing Necessary Condition 4 in polynomial time.
  
\end{unsolved}

%\subsubsection{Visibility Graph Characterization}
\subsection{Visibility Graph Characterization}

The problem of identifying the set of properties satisfied by all 
visibility graphs of simple polygons
is called the  visibility graph 
{\it characterization} problem for polygons.
Let us state some results on the problems of characterizing visibility graphs for special classes of 
simple polygons. The earliest result is from ElGindy \cite{el-hcpa-85} who showed that every 
{\it maximal outerplanar graph} is a visibility graph of a simple polygon, and he suggested 
an $O(n \log n)$ algorithm for reconstruction. If all reflex vertices of a simple polygon occur 
consecutively along its boundary, the polygon is called a {\it spiral polygon}. Everett and Corneil 
\cite{ev-vgc-90,ec-rvgsp-90} characterized visibility graphs of spiral polygons by showing that these 
graphs are a subset of {\it interval graphs} which lead to an $O(n)$  time algorithm.  
Choi, Shin and Chwa \cite{ChoiSC95} characterized funnel-shaped polygons, also called 
{\it towers}, and gave an $O(n)$ time recognition algorithm. Visibility graphs of towers are also 
characterized by Colley, Lubiw and Spinrad \cite{cls-vgt-97} and they have shown that visibility graphs 
of towers are bipartite permutation graphs with an added Hamiltonian cycle. If the internal angle at 
each vertex of a simple polygon is either 90 or 270 degrees, then the polygon is called a 
{\it rectilinear polygon}. If the boundary of a rectilinear polygon is formed by a staircase path with 
two other edges, the polygon is called a {\it staircase polygon}. Visibility graphs of staircase polygons 
have been characterized by Abello, E\~{g}ecio\~{g}lu, and Kumar \cite{AbeEgeKum95}. Lin and Chen 
\cite{lc-pvg-94} have studied  visibility graphs that are {\it planar}.

For the characterization of visibility graphs of arbitrary simple polygons, 
Ghosh has shown that visibility graphs do not possess the characteristics of {\it perfect graphs}, 
{\it circle graphs} or {\it chordal graphs}. On the other hand, Coullard and Lubiw \cite{cl-dig-92} 
have proved that every triconnected component of a visibility graph satisfies {\it 3-clique ordering}. 
This property suggests that structural properties of visibility graphs may be related to well-studied 
graph classes, such as {\it 3-trees} and {\it 3-connected graphs}. Everett and Corneil 
\cite{ev-vgc-90,ec-nrcvg-95} have shown that there is no finite set of forbidden induced subgraphs that 
characterize visibility graphs. Abello and Kumar \cite{ak-lncs-1995, Abello:1994:VGO} have suggested a 
set of necessary conditions for recognizing visibility graphs. However, it has been 
shown in \cite{g-rcvg-97} 
that this set of conditions follow from the last two necessary conditions of Ghosh.
For more details on the characterization of visibility graphs, see Ghosh \cite{g-vap-07}.
\begin{unsolved}
Characterize visibility graphs of simple polygons.
\end{unsolved}

%\subsubsection{Visibility Graph Reconstruction}
\subsection{Visibility Graph Reconstruction}

The problem of actually drawing a simple polygon $P$ whose 
visibility graph is the given graph $G$, is called the visibility graph {\it reconstruction} problem
for polygons. 
Let us mention some of the approaches on the visibility graph reconstruction problem. It has been shown 
by Everett \cite{ev-vgc-90} that visibility graph reconstruction is in {\it PSPACE}. This is the only 
upper bound known on the complexity of the problem. Abello and Kumar \cite{Abello:1994:VGO} studied the 
relationship between visibility graphs and oriented matroids, Lin and Skiena \cite{ls-cavg-95} studied 
the equivalent order types, and Streinu \cite{is-cccg-1999,is-cgta-2003} and O'Rourke and Streinu 
\cite{os-vpg-97} studied psuedo-line arrangements. Everett and Corneil \cite{ev-vgc-90,ec-nrcvg-95} 
have solved the reconstruction problem for the visibility graphs of \emph{spiral} polygons and the
corresponding problem for the visibility graph of \emph{tower} polygons has been solved by Choi, Shin 
and Chwa \cite{ChoiSC95}. Reconstruction problem with added information has been studied by Coullard 
and Lubiw \cite{cl-dig-92},  Disser,  Mihal\'{a}k, and Widmayer \cite{dmw-rspa-2010},
Everett, Hurtado, and Noy \cite{ehn-dam-1999}, Everett, Lubiw, and O'Rourke 
\cite{elo-cccg-1993}, Jackson and Wismath \cite {jw-cgta-2002}.

\begin{unsolved}
Draw a simple polygon whose visibility graph is the given graph $G$.
\end{unsolved}

%\subsection{Graph Theoretic Problems on Visibility Graphs}

%\subsubsection{Hamiltonian Cycle in Visibility Graphs}
\subsection{Hamiltonian Cycle in Visibility Graphs}

A \emph{Hamiltonian cycle} is a cycle in an undirected graph which visits each vertex exactly once 
and also returns to the starting vertex. The Hamiltonian cycle problem is to determine whether a 
Hamiltonian cycle exists in a given graph $G$. Observe that $G$ may contain 
several Hamiltonian cycles, and $G$ may be visibility graph for a Hamiltonian cycle and is 
not a valid visibility graph for another Hamiltonian cycle in $G$. 

\begin{unsolved}
Given the visibility graph $G$ of a simple polygon $P$, determine the Hamiltonian cycle in $G$ 
that corresponds to the boundary of $P$.
\end{unsolved}

%\subsubsection{Minimum Dominating Set in Visibility Graphs}
\subsection{Minimum Dominating Set in Visibility Graphs}

A \emph{dominating set} for a graph $G = (V, E)$ is a subset $D$ of $V$ such that every vertex 
not in $D$ is joined to at least one member of $D$ by some edge. The minimum dominating 
set problem in visibility graphs corresponds to the art gallery problem in polygons which has been 
shown to be NP-hard \cite{ll-ccagp-86, ls-cavg-95}. Following the approximation algorithm for the 
art gallery problem for polygons given by Ghosh \cite{g-apat-2006,g-apatpp-2010}, 
a minimum dominating set of 
visibility graph can be computed with an approximation ratio of $O(\log n)$.

\begin{unsolved}
 Design a constant factor approximation algorithm for computing minimum dominating set  of 
visibility graphs.
\end{unsolved}

%\subsubsection{Maximum Hidden Set in Visibility Graphs}
\subsection{Maximum Hidden Set in Visibility Graphs}

An \emph{independent set} is a set of vertices in a graph with no two of which are adjacent. Independent 
sets in visibility graphs are known as {\it hidden vertex sets}. Shermer \cite{s-hpip-89} has proved 
that the maximum hidden vertex set problem on visibility graphs is also NP-hard. However, the problem 
may not remain NP-hard if the Hamiltonian cycle corresponding to the boundary of the simple polygon 
is given as an input along with the visibility graph. With this additional input, Ghosh, Shermer, 
Bhattacharya and Goswami \cite{gsbg-cmcvgsp-07} have shown that it is possible to compute in $O(ne)$ 
time the maximum hidden vertex set  in the visibility graph of a very special class of simple polygons 
called {\it convex fans}, where $n$ and $e$ are the number of vertices and edges of the input 
visibility graph of the convex fan respectively. Hidden vertex sets are also studied by Eidenbenz 
\cite{se-eth-2000,se-cgta-2002}, Ghosh et al. \cite{gmpsv-crwvp-93} and Lin and Skiena 
\cite{ls-cavg-95}. 

\begin{unsolved}
 Given the visibility graph $G$ of a simple polygon $P$ along with the Hamiltonian cycle in $G$ 
corresponding to the boundary of $P$, determine the maximum hidden set of $G$.
\end{unsolved}

\begin{unsolved}
Design an approximation algorithm for computing maximum hidden set of visibility graph.
\end{unsolved}

%\subsubsection{Maximum Clique in Visibility Graphs}
\subsection{Maximum Clique in Visibility Graphs}

A \emph{clique} in a graph is a set of pairwise adjacent vertices. The problem of computing the maximum 
clique in the visibility graph is not known to be NP-hard. Observe that the maximum clique in a 
visibility graph corresponds to the largest empty convex polygon inside the corresponding polygon. 
Algorithms for computing largest empty convex polygons has been reported by several authors 
\cite{ar-clec-85, deo-secp-90, es-mcmcpvg-00}. However, for each of these algorithms, the input is 
either a polygon \cite{es-mcmcpvg-00} or a point set \cite{ar-clec-85, deo-secp-90}. Spinrad 
\cite{s-grep-03} has discussed possible approaches for computing maximum clique using the notion 
of \emph{triangle-extendible ordering} which is essentially a transitive orientation of the graph.

\begin{unsolved}
 Given a visibility graph $G$ in adjacency matrix form, compute a maximum clique of $G$.
\end{unsolved}

\begin{unsolved}
 Determine whether a set of vertices of a visibility graph has a triangle-extendible ordering in 
polynomial time.
\end{unsolved}

If the Hamiltonian cycle in a visibility graph corresponding to the boundary of the polygon 
is given along with the visibility graph as an input, Ghosh, Shermer, Bhattacharya and Goswami 
\cite{gsbg-cmcvgsp-07} have presented an $O(n^2e)$ time algorithm for computing the maximum 
clique in the visibility graph $G$ of a simple polygon $P$. Here $n$ and $e$ are number of 
vertices and edges of $G$ respectively.

\subsection{Counting Visibility Graphs}
Two graphs $G_1 = (V_1, E_1)$ and $G_2 = (V_2, E_2)$ are \emph{isomorphic} if and only if there 
is a bijection $f$ that maps vertices of $V_1$ to the vertices of $V_2$ such that an edge 
$(v, w) \in E_1$ if and only if the edge $(f(v), f(w)) \in E_2$. It has been shown 
by Ghosh \cite{g-vap-07} that the number of \emph{non-isomorphic} visibility graphs of 
simple polygons of $n$ vertices is 
at least $2^{n-4}$. On the other hand, a straightforward application of Warren's theorem 
\cite{w-lbanm-68} shows that the number of visibility graphs is at most $2^{O(n\log n)}$ 
\cite{s-grep-03}.

\begin{unsolved}
Improve the lower and upper bounds on the number of {\it non-isomorphic} visibility graphs of 
simple polygons.
\end{unsolved}

Let $G_1$ and $G_2$ be the visibility graphs of simple polygons $P_1$ and $P_2$ respectively. Let
$C_1$ (or $C_2$) denote the Hamiltonian cycles in $G_1$ (respectively, $G_2$) that corresponds to 
the boundary of $P_1$ (respectively, $P_2$). Polygons $P_1$ and $P_2$ are called \emph{similar} if
and only if there is a bijection $f$ that maps adjacent vertices on the boundary of $P_1$ to that of
boundary of $P_2$ such that $f(G_1) = G_2$ \cite{ls-cavg-95}. It has been shown \cite{ae-caps-83} 
that similarity of $P_1$ and $P_2$, each of $n$ vertices, can be determined in $O(n^2)$ time. 
Therefore, given $G_1$ and $G_2$ along with $C_1$ and $C_2$, the corresponding visibility graph 
similarity problem can also be solved in $O(n^2)$ time. It has been shown by Lin and Slkiena 
\cite{ls-cavg-95} that two simple polygons with isomorphic visibility graphs may not be similar 
polygons.

\subsection{Representing Visibility Graphs}
Although the most natural form of representation for visibility graphs would be to use coordinates 
of the points, this is not useful if we are looking for a space efficient representation. Lin and 
Skiena \cite{ls-cavg-95} have proved that visibility graphs require endpoints to have 
exponential sized integers. However, it is not known whether singly exponential sized integers are 
sufficient. It is important because if we could guarantee that the number of bits in the integer is 
polynomial, then visibility graph recognition is in $NP$ \cite{s-grep-03}.

\begin{unsolved}
Can all endpoints of a visibility graph be assigned integer coordinates such that the integers use 
a polynomial number of bits?
\end{unsolved}

A natural form of storage is studied by Agarwal et al.\ \cite{aaas-cvgr-94} which uses a relatively 
small number of bits to store a visibility graph. However, the representation is neither space 
optimal, nor adjacency information can be retrieved in constant time. However, it is the most 
significant reduced space representation which is currently known. The authors consider the problem 
of representing 
a visibility graph as a covering set of cliques and complete bipartite graphs so that every graph 
in the set is a subset of $G$, and every edge is contained in at least one of the graphs of the 
covering set. Their proposed algorithm constructs a covering set which has $O(n\log^4 n)$ bits. It 
can be shown that any covering set requires $\Omega(n \log^2 n)$ bits on some visibility graphs 
\cite{s-grep-03}. Given this gap between upper and lower bounds on this natural form of 
representation, we have a number of problems.

\begin{unsolved}
 Give a tight bound (with respect to order notation) on the number of bits used in an optimal 
covering set of a visibility graph.
\end{unsolved}

\begin{unsolved}
 Find a covering set which matches the above bound in polynomial time.
\end{unsolved}

\bibliographystyle{plain}
\bibliography{vis}

\end{document}